\documentclass[final]{article}
\usepackage[final]{graphicx}
\usepackage{siunitx}
\usepackage{graphicx}
\usepackage{amssymb}
\usepackage{rotating}
\usepackage{pdflscape}
\usepackage[english]{babel}
\usepackage{color}
\usepackage{bigintcalc}
\def\beq{\begin{eqnarray}}
\def\eeq{\end{eqnarray}}
\usepackage{amsfonts}

\usepackage{amsmath}
\usepackage{bbm}
\begin{document}


\title{Quantum particles in a suddenly accelerating potential}

\author{
Paolo Amore \\
Facultad de Ciencias, CUICBAS, Universidad de Colima,\\
Bernal D\'{i}az del Castillo 340, Colima, Colima, Mexico  \\
paolo.amore@gmail.com \\
Francisco M. Fern\'andez \\
INIFTA, Divisi\'{o}n Qu\'{\i}mica Te\'{o}rica,\\
Blvd. 113 y 64 (S/N), Sucursal 4, Casilla de Correo 16,\\
1900 La Plata, Argentina \\
Jos\'e Luis Valdez \\
Facultad de Ciencias, Universidad de Colima,\\
Bernal D\'{i}az del Castillo 340, Colima, Colima, Mexico  }

\maketitle

\begin{abstract}
We study the behavior of a quantum particle trapped in a confining potential in one dimension under multiple sudden changes of velocity  and/or acceleration. 
We develop the appropriate formalism to deal with such situation and we  use it to calculate the probability of 
transition for simple problems such as the particle in an infinite box and the simple harmonic oscillator. For the infinite box of length $L$ under two and three sudden changes of velocity, where the initial and final velocity vanish, we find that the system undergoes quantum revivals for $\Delta t = \tau_0 \equiv \frac{4mL^2}{\pi\hbar}$, regardless of other parameters ($\Delta t$ is the time elapsed between the first and last change of velocity).   For the simple harmonic oscillator 
we find that the states obtained by suddenly changing (one change) the velocity and/or the acceleration of the potential, for a particle initially in an eigenstate
of the static potential,  are {\sl coherent} states. For multiple changes of acceleration or velocity we find that the quantum expectation value of the Hamiltonian is remarkably close (possibly identical) to the corresponding classical expectation values. Finally, the probability of transition for a particle in an accelerating 
harmonic oscillator (no sudden changes) calculated with our formalism agrees with the formula derived long time ago by Ludwig~\cite{Ludwig51} and recently modified by Dodonov~\cite{Dodonov21}, but with a different expression for the dimensionless parameter $\gamma$. Our probability agrees with the one of ref.~\cite{Dodonov21} for $\gamma \ll 1$ but is not periodic in time (it decays monotonously), contrary to the result derived in ref.~\cite{Dodonov21}.
\end{abstract}

\section{Introduction}
\label{Intro}

In this paper we consider the  quantum mechanical problem of a non relativistic particle in a potential in one dimension, that can undergo multiple sudden changes
of both velocity and acceleration at given times. 

This problem falls in the larger class of problems where the Hamiltonian of the system displays an explicit dependence on the time. In the most common situation this explicit time dependence is the result of a time--dependent interaction  or of a time--independent  interaction that acts over a finite  interval of time; alternatively, the time-dependence can enter in the problem through time--dependent boundary conditions or through a not stationary potential  (for example, if the potential suddenly changes velocity).

An important example of problem belonging to the first class is the infinite square well with a moving wall, studied long time ago by Doescher and Rice~\cite{Doescher69}, for which the solution can be calculated exactly (a generalization of this problem for a wall moving with different laws 
has been obtained in \cite{Makowski91}).
In the second class of problems we mention  the case of a particle in a one dimensional attractive Dirac delta potential, that suddenly starts to move, originally studied by  Granot and Marchewka ~\cite{Granot09}. This problem has been recently revisited in a paper by two of the present authors~\cite{Amore19}, extending the analysis to different potentials, using an approach based on a direct spectral decomposition. Additionally, 
a recent paper by Dodonov~\cite{Dodonov21} considers the interesting problem of calculating the transition amplitudes for a quantum harmonic oscillator (in one dimension) moving with an arbitrary law of motion (this paper also offers a rich bibliography on the subject, that could be of interest to the reader).

In this paper we consider the more general problem of calculating the probability of transition for a quantum particle in one dimension when the potential is  subject to an arbitrary number of sudden changes in the velocity and/or the acceleration. This discussion constitutes a generalization of our previous work  \cite{Amore19}, which only considered one change in velocity. 

The paper is organized as follows: in section  \ref{sec:1} we discuss the solution to the nonrelativistic Schr\"odinger equation in presence of acceleration; in section \ref{sec:2} we describe the suitable formalism to deal with sudden changes of the velocity and the acceleration (for technical reasons the discussion is done separately); in section \ref{sec:3} we apply our formalism to study a infinite square well under multiple changes of velocity and a simple harmonic oscillator under multiple changes  of both velocity and acceleration (in both cases,  we compare the quantum results with the analogous classical results). Finally, in section \ref{sec:4} we draw our conclusion and outline the directions of future work.

\section{Accelerating potential}
\label{sec:1}

We consider the time--dependent Schr\"odinger equation for a potential undergoing a constant acceleration
\begin{eqnarray}
i \hbar \frac{\partial \psi(x,t)}{\partial t} = - \frac{\hbar^2}{2m} \frac{\partial^2 \psi}{\partial x^2} + 
V\left(x-x_0-v (t-T) - \frac{1}{2} a (t-T)^2 \right) \psi(x,t) \ .
\label{eq_Sch}
\end{eqnarray}

Notice that for $a=0$ one recovers the case studied in Ref.~\cite{Amore19} of a potential  moving with velocity $v$. 

For simplicity we define $\xi(x,t) \equiv x-x_0-v (t-T) - \frac{1}{2} a (t-T)^2$ and we look for a solution of the form
\begin{eqnarray}
\psi(x,t) = \phi\left(\xi(x,t) \right)  \ e^{i \sigma(x,t)} \ .
\label{eq_ansatz}
\end{eqnarray}

Upon substitution of this equation inside eq.~(\ref{eq_Sch}) one obtains the equation
\begin{equation}
\begin{split}
\frac{\hbar^2}{2 m} \frac{d^2\phi}{d\xi^2}    
&-\frac{i \hbar}{m}  \left(m (a (t-T)+v)-\hbar  \frac{\partial\sigma}{\partial x}\right) \frac{d\phi}{d\xi} \\
& + \left( -V(\xi)  - \hbar \frac{\partial \sigma}{\partial t}  - \frac{\hbar^2}{2m} \left(\frac{\partial \sigma}{\partial x}\right)^2 + i
\frac{\hbar^2}{2m} \frac{\partial^2 \sigma}{\partial x^2} \right) \phi(\xi)
 = 0 \ .
\end{split}
\label{eq_sub}
\end{equation}

To be able to interpret eq.~(\ref{eq_sub}) as a time--independent Schr\"odinger equation we need to eliminate the  term proportional to $d\phi/d\xi$ using the condition 
\begin{equation}
\frac{\partial\sigma}{\partial x} = \frac{m}{\hbar} (a (t-T)+v) .
\end{equation}

The solution to this differential equation is easily obtained as 
\begin{equation}
\sigma(x,t) = \frac{m x (a (t-T)+v)}{\hbar }+Q(t) \ , 
\end{equation}
where $Q(t)$ is a function only of time; upon substitution of this solution inside eq.~(\ref{eq_sub}) we obtain 
\begin{equation}
\begin{split}
\frac{\hbar ^2}{2 m}  \frac{d^2\phi}{d\xi^2} & + \left(-a^2 m (t-T)^2-a m \xi -2 a m v (t-T) \right. \\
&\left. -a m x_0-\frac{m v^2}{2}-\hbar  \dot{Q}(t)-V(\xi ) \right) \phi(\xi)  = 0  \ .
\label{eq_tise_1}
\end{split}
\end{equation}
$Q(t)$ is determined by asking that the term multiplying $\phi \left(\xi\right)$ in this equation 
be a function of $\xi(x,t)$ only.  By enforcing this condition we find
\begin{equation}
\begin{split}
Q(t) &= -\frac{a^2 m (t-T)^3}{3 \hbar} -\frac{a m v (t-T)^2}{\hbar}
-\frac{a m x_0 (t-T)}{\hbar} -\frac{m v^2 (t-T)}{2 \hbar}+\frac{\epsilon  (T-t)}{\hbar }  \ ,
\end{split}
\end{equation}
where $\epsilon$ is a constant with units of energy; upon substitution inside eq.~(\ref{eq_tise_1}) we obtain 
\begin{equation}
-\frac{\hbar ^2}{2 m} \frac{d^2\phi(\xi)}{d\xi^2} + \left(a m \xi +V(\xi) \right) \phi(\xi)= \epsilon \phi(\xi) \ ,  
\label{eq_tise_2}
\end{equation}
i.e. the time--independent Schr\"odinger equation in the {\sl moving} coordinate, in presence of the original 
potential and of a extra term $m a \xi$~\footnote{The case of the simple harmonic oscillator (sho) is peculiar because under 
a constant acceleration the potential transforms into a displaced sho.}. 

The full time--dependent solution will then be
\begin{equation}
\Psi(x,t) = \phi (\xi )  \exp \left(i \sigma(x,t)\right)  \ .
\label{eq_psi_va}
\end{equation}
where
\begin{equation}
\begin{split}
\sigma(x,t) &= \frac{a^2 m (t-T)^3}{6 \hbar }+\frac{a m (t-T) \xi}{\hbar }+\frac{a m v (t-T)^2}{2 \hbar} \\
&+\frac{m v^2 (t-T)}{2 \hbar }+\frac{m \xi v}{\hbar }+\frac{m v x_0}{\hbar}-\frac{\epsilon  (t-T)}{\hbar } \ .
\end{split}
\nonumber
\end{equation}

Notice that for $a=0$ this expression reduces to
\begin{equation}
\begin{split}
\sigma(x,t) = \frac{m v^2 (t-T)}{2 \hbar }+\frac{m \xi v}{\hbar }+\frac{m v x_0}{\hbar}-\frac{\epsilon  (t-T)}{\hbar } \ ,
\end{split} 
\nonumber
\end{equation}
previously studied in ref.~\cite{Amore19}.

The  expression that we have described is a special case of the transformations originally obtained by Butkovskiy and Samoilenko~\cite{Butkovskiy84},  which allow one to deal with  more general motion of the potential. Notice however that 
the Schr\"odinger equation in the rest frame of the potential  adquires an explicit time dependence unless the acceleration and the velocity
are constant (see eqs.(1) and (2) of ref.~\cite{Rouchon03}).

The reader interested in a deeper discussion on the Schrodinger equation in accelerating frames may  see  refs.~\cite{Robinett96, Klink97, Vandergrift00, Torres17}.

\section{Multiple changes of velocity and/or acceleration}
\label{sec:2}

In this section we consider a potential which can suddenly change its velocity and/or acceleration at given times. 
In particular, we want to describe the behavior of a quantum particle trapped in this potential, as the potential suddenly moves.

For technical reasons we will discuss separately the case where only changes in velocity occur and the case where both 
acceleration and velocity can change.

\subsection{Sudden changes of velocity}

In this case,  the potential in eq.~(\ref{eq_tise_2}) does not change form and the discussion can be easily done for arbitrary (confining or not) potentials.  

We assume that the potential $V(x)$  changes velocity at selected times $t_1$, $t_2$, \dots ($t_1 < t_2 < \dots$).
At the time $t_i$ the velocity of the potential suddenly changes from $v_i$ to $v_{i+1}$; the time--dependent potential
in this case reads
\begin{equation}
\begin{split}
\mathcal{V}(x,t) =  \left\{
\begin{array}{ccc}
V(x-x_1-v_1 (t-t_1)) & , & t<t_1 \\
V(x-x_2-v_2 (t-t_2)) & , & t_1<t<t_2  \\
V(x-x_3-v_3 (t-t_3)) & , & t_2<t<t_3  \\ 
&\dots \ , 
\end{array}
\right. 
\end{split}
\label{eq_pot_v}
\end{equation}
where
\begin{equation}
\begin{split}
x_k &= x_1 + \sum_{j=2}^k v_j (t_j-t_{j-1}) \hspace{1cm} , \hspace{1cm} k=2,3,\dots \  .
\end{split}
\nonumber
\end{equation}

The above quantities can be expressed in terms of the moving coordinates
\begin{equation}
\begin{split}
\xi_k(x, t) &\equiv x - x_k -v_k (t-t_k)  \ .
\end{split}
\nonumber
\end{equation}

In each time interval $t_i \leq t \leq t_{i+1}$ one can decompose the wave function describing the particle 
in the basis of the moving potential  using  eq.~(\ref{eq_psi_va}) setting $a=0$.  These eigenfunctions take the form
\begin{equation}
\begin{split}
\psi_k(x,t;X,T;v)  &= \psi_k (x-X-v (t-T)) \ e^{-i \frac{m v^2 (t-T)}{2\hbar} + i \frac{m v (x-X) }{\hbar} - i \frac{E_k (t-T)}{\hbar}}
\end{split}
\label{eq_psik}
\end{equation}
where  $\psi_k(x)$ are the eigenfunctions of the static potential $V(x)$ (notice that 
$\psi_k(x,T;0,T;0) = \psi_k(x)$). The quantum number $k$ can be either discrete or continuum depending on the nature of the potential 
(for simplicity we will restrict to confining potentials and therefore deal only with discrete values).

The expectation value of momentum and position for a particle described by this wave function are easily obtained as
\begin{equation}
\begin{split}
\langle p  \rangle &\equiv \int \psi^\star_k(x,t;X,T;v) \ \hat{p} \ \psi_k(x,t;X,T;v) dx  = \langle \hat{p} \rangle_{v=0} + m v \\
\langle x  \rangle &\equiv \int \psi^\star_k(x,t;X,T;v) \ x \ \psi_k(x,t;X,T;v) dx  = \langle \hat{x} \rangle_{X=0,v=0} + X + v (t-T)  \ ,
\end{split}
\nonumber
\end{equation}
where $\langle \hat{O} \rangle_{v_0}$ denotes the expectation value in the static potential, whereas the remaining terms 
are associated to the movement of the potential.

It is important to remark that the set $\left\{ \psi_k(x,t;X,T;v)\right\}$ forms a basis for the Hamiltonian of a 
potential moving with velocity $v$ (in other words, each choice of the parameters $X$, $T$ and $v$ will define a particular basis).
The wave function in each time interval will be more conveniently described by the basis for the potential moving with appropriate
velocity and with the proper initial conditions ($X$ defines the position of the potential at time $T$).

We will write the wave function as
\begin{equation}
\Psi(x,t) = \left\{ \begin{array}{ccc}
\Psi^{(1)}(x,t) & , & t \leq t_1 \\
\Psi^{(2)}(x,t) & , & t_1\leq  t \leq  t_2 \\
\Psi^{(3)}(x,t) & , & t_2\leq  t \leq  t_3 \\
\dots & & 
\end{array} \right.  \  , \nonumber
\end{equation}
and assume that it is known at a given time, for example, at $t=t_1$. The time evolution of this wave function will be governed by
the corresponding time--dependent Schr\"odinger equation, requiring that it is continuous at each time where a change of velocity occurs~\footnote{Clearly the TDSE is first order in time and therefore a single condition is sufficient.}, i.e.
\begin{equation}
\begin{split}
\Psi^{(1)}(x,t_1) &= \Psi^{(2)}(x,t_1)  \\
\Psi^{(2)}(x,t_2) &= \Psi^{(3)}(x,t_2)  \\
&\dots   \ .
\end{split}
\label{eq_match}
\end{equation}

The time evolution of a wave packet for the case in which there is no change of velocity does not require specifying any matching condition (this is the case usually 
discussed in Quantum Mechanics textbooks).  

Once the solution is known in a temporal region $t_i \leq t  \leq t_{i+1}$, the solution in the neighboring temporal regions,
 $t_{i-1} \leq t  \leq t_{i}$ or  $t_{i+1} \leq t  \leq t_{i+2}$, can be obtained imposing the equations (\ref{eq_match}).

We will now discuss in detail the procedure to obtain the solution at different times and in different temporal regions.

\begin{itemize}
\item $t< t_1$;
    
The "initial wave function" can be decomposed in terms of the basis moving with velocity $v_1$ for a potential initially centered at $x_1$ at time $t_1$:
\begin{equation}
\begin{split}
\Psi^{(1)}(x,t) &= \sum_l c_l^{(1)} \psi_l(x,t;x_1,t_1,v_1)  \\
&= \sum_l c_l^{(1)} \ \psi_l(x-x_1 - v_1 (t-t_1))  \ e^{-i \frac{m v_1^2 (t-t_1)}{2\hbar} + i \frac{m v_1 (x-x_1) }{\hbar} - i \frac{E_l (t-t_1)}{\hbar}}  \ . 
\end{split}
\label{eq_t1_a}
\end{equation}

The coefficients of this expansion are easily obtained by multiplying both sides of the equation
above at $t=t_1$ by the factor $ \psi_l^\star(x-x_1) \ e^{- i \frac{m v_1 (x-x_1) }{\hbar} }$ and integrating over space
\begin{equation}
\begin{split}
c_l^{(1)} &= \int_{-\infty}^{\infty} \psi_l^\star(x-x_1) \Psi^{(1)}(x,t_1) e^{- i \frac{m v_1 (x-x_1) }{\hbar} } \ dx   \ .
\end{split}
\label{eq_t1_b}
\end{equation}

By using (\ref{eq_t1_b}) inside eq.~(\ref{eq_t1_a}) we have
\begin{equation}
\begin{split}
\Psi^{(1)}(x,t) &= e^{-i \frac{m v_1^2 (t-t_1)}{2\hbar}} \\
&\cdot \int_{-\infty}^{\infty}  e^{ i \frac{m v_1 (x-y) }{\hbar} }  \ K(x-x_1 - v_1 (t-t_1),t;y-x_1,t_1) \ \Psi^{(1)}(y,t_1)   \ dy  \  , 
\end{split}
\label{eq_t1_c}
\end{equation}
where  $K$ is  the quantum propagator defined as
\begin{equation}
K(x,t;y,t_0) = \sum_j \psi_j(x) \psi_j^\star(y) e^{-i E_j (t-t_0)/\hbar}  \ .
\nonumber 
\end{equation}

Notice that the explicit expression of $K$ is known only in special cases, such as  for the  simple harmonic oscillator in one dimension,
where it takes the form
\begin{equation}
\begin{split}
K(x,t;y,t_0) &= \sqrt{\frac{m \omega}{2\pi i \hbar \sin(\omega(t-t_0))}} \  e^{\frac{i m \omega}{2 \hbar \sin(\omega(t-t_0))} \left[ (x^2 +y^2) \cos(\omega (t-t_0)) -2 x y\right]}  \ . 
\nonumber
\end{split}
\end{equation}

In particular for $v_1=0$ and $x_1=0$ eq.~(\ref{eq_t1_c}) reduces to the standard formula 
\begin{equation}
\begin{split}
\Psi^{(1)}(x,t) &= \int_{-\infty}^{\infty}  K(x,t;y,t_1) \ \Psi^{(1)}(y,t_1)   \ dy \ .
\end{split}
\label{eq_t1_d}
\end{equation}

\item $t_1 < t< t_2$;

As we have already observed, a basis for a potential moving with velocity $v$ is fully determined only when the parameters $X$ and $T$
are specified.  In the case of a potential moving with velocity $v_2$ for $t_1 \leq t \leq t_2$, there are two natural choices of basis:
$\left\{ \psi_k(x,t;x_1,t_1;v_2)\right\}$ and $\left\{ \psi_k(x,t;x_2,t_2;v_2)\right\}$. In the first case the basis relates to a potential that at time $t_1$ is centered at $x_1$ and moves with velocity $v_2$, whereas in the second case the basis relates to a potential that at time $t_2$ is centered at $x_2$ and moves with velocity $v_2$. What makes these bases special, among the infinite bases available, is the fact that they allow one to enforce the matching conditions at the times $t=t_1$ and $t=t_2$, where the potential undergoes a sudden change of velocity.

We can decompose the wave function $\Psi^{(2)}(x,t)$ in any of the two bases; in the first case we have
\begin{equation}
\begin{split}
\Psi^{(2)}(x,t) &= \sum_l c_l^{(2)} \psi_l(x,t;x_1,t_1,v_2)   \ ,
\label{eq_first}
\end{split}
\end{equation}
while in the second case we have
\begin{equation}
\begin{split}
\Psi^{(2)}(x,t) &= \sum_l \bar{c}_l^{(2)} \psi_l(x,t;x_2,t_2,v_2)   \ .
\label{eq_second}
\end{split}
\end{equation}

By comparing eqs.~(\ref{eq_first}) and (\ref{eq_second}) we are able to relate the coefficients of the expansion in the two basis:
\begin{equation}
c_l^{(2)} = \bar{c}_l^{(2)} \ e^{-i \frac{m v_2^2 (t_1-t_2)}{2\hbar}} \  e^{i \frac{m v_2  (x_1-x_2) }{\hbar} } \ e^{- i \frac{E_l (t_1-t_2)}{\hbar}}  \ .
\end{equation}

It is important to remark that, for the case of a sudden change of velocity, the coefficients in the two basis differ only by a phase (we will soon see that this
relation is considerably more complicated in presence of a sudden acceleration).

The next step is to enforce the matching condition $\Psi^{(1)}(x,t_1) = \Psi^{(2)}(x,t_1)$
obtaining
\begin{equation}
\begin{split}
c_k^{(2)} &=  \sum_{l} \mathcal{Q}_{kl}(v_1,v_2;t_1) c_l^{(1)} ,
\end{split}
\label{eq_c2}
\end{equation}
where
\begin{equation}
\begin{split}
\mathcal{Q}_{kl}(v_1,v_2;t_1) &\equiv \int_{-\infty}^\infty \psi_k^\star(y)  \psi_l(y) \ e^{-i \frac{m (v_2-v_1) y}{\hbar}} \ dy  .
\end{split}
\nonumber
\end{equation}

It is straightforward to check that $\mathcal{Q}$ is unitary and therefore eq.~(\ref{eq_c2}) complies with the conservation of probability 
\begin{equation}
\begin{split}
\sum_k |c_k^{(2)}|^2  &= \sum_k |c_k^{(1)}|^2 = 1 \ .
\end{split}
\end{equation}

Now we are in position of expressing the wave function for $t_1 < t < t_2$ in terms of the initial wave function as
\begin{equation}
\begin{split}
\Psi^{(2)}(x,t) &= \sum_k c_k^{(2)} \psi_k(x,t;x_1,t_1,v_2)  \\
&= e^{-\frac{i m v_2^2 (t-t_1)}{2\hbar }+\frac{i m  v_2 (x-x_1)}{\hbar}}  \sum_l  c_l^{(1)} \int_{-\infty}^\infty e^{-i \frac{m (v_2-v_1) y}{\hbar}} K(x-x_1-v_2 (t-t_1),t; y,t_1)   \psi_l(y) dy  \ ,
\label{eq_psi}
\end{split}
\end{equation}
where
\begin{equation}
 K(x-x_1-v_2 (t-t_1),t; y,t_1)  = \sum_k e^{\frac{-i E_k (t-t_1)}{\hbar }} \ \psi_k(x-x_1 -v_2 (t-t_1)) \ \psi_k^\star(y)  \ .
 \nonumber 
\end{equation}

Eq.~(\ref{eq_psi}) can finally be cast in a more convenient form 
\begin{equation}
\begin{split}
\Psi^{(2)}(x,t) &= e^{-\frac{i m v_2^2 (t-t_1)}{2\hbar }+\frac{i m  v_2 (x-x_1)}{\hbar}}   e^{-i \frac{m v_1 x_1}{\hbar}}   \int_{-\infty}^\infty e^{-i \frac{m v_2 y}{\hbar}} \  K(x-x_1-v_2 (t-t_1),t; y,t_1) \Psi^{(1)}(y,t_1) dy  \ , 
\label{eq_psib}
\end{split}
\end{equation}
which reduces, once again, to eq.~(\ref{eq_t1_d}) for $v_1=v_2=0$.

For  $t_{j-1} < t < t_j$ we can generalize this formula to
\begin{equation}
\begin{split}
\Psi^{(j)}(x,t) &= e^{-\frac{i m v_j^2 (t-t_{j-1})}{2\hbar }+\frac{i m  v_j (x-x_{j-1})}{\hbar}}   
e^{-i \frac{m v_{j-1} x_{j-1}}{\hbar}}  \\
&\cdot \int_{-\infty}^\infty e^{-i \frac{m v_j y}{\hbar}} \  K(x-x_{j-1}-v_j (t-t_{j-1}),t; y,t_{j-1}) \Psi^{(j-1)}(y,t_{j-1}) dy   \ .
\label{eq_psi_j}
\nonumber
\end{split}
\end{equation}

\end{itemize}

\subsection{Sudden changes of velocity and acceleration}

We will now consider the case in which the potential undergoes sudden changes of the acceleration (and possibly of the velocity)
at specific times. Although the discussion can be made for general potentials, as done in the previous section, we will concentrate
on the simple harmonic oscillator. The reason for this choice is the unique nature of the SHO, which transforms into another SHO  under a uniform acceleration: having to deal only with SHOs will make the technical part of our discussion simpler and allow to  avoid to a large extent numerical approximations~\footnote{If one chooses, for example, a particle in an infinite box, under a constant acceleration the potential behaves linearly between the walls and the spectrum as well as the wave functions have to be determined numerically. }.

In this case the potential is
\begin{equation}
\begin{split}
\mathcal{V}(x,t) =  \left\{
\begin{array}{ccc}
\frac{1}{2} m \omega^2 (x-x_1-v_1 (t-t_1)-\frac{1}{2} a_1 (t-t_1)^2)^2 & , & t \leq t_1 \\
\frac{1}{2} m \omega^2 (x-x_2-v_2 (t-t_2)-\frac{1}{2} a_2 (t-t_2)^2)^2 & , & t_1\leq t\leq t_2  \\
\frac{1}{2} m \omega^2 (x-x_3-v_3 (t-t_3)-\frac{1}{2} a_3 (t-t_3)^2)^2 & , & t_2\leq t\leq t_3  \\
&\dots \ ,
\end{array}
\right. 
\end{split}
\label{eq_pot_a}
\end{equation}
where $x_k = x_1 + \sum_{j=2}^k v_j (t_j-t_{j-1}) + \frac{1}{2} \sum_{j=2}^k a_j (t_j -t_{j-1})^2$ ($k=2,3,\dots$).

We work in the moving coordinate system represented by
\begin{equation}
\xi_j(x,t) =  x - x_j - v_j (t-t_j) - \frac{1}{2} a_j (t-t_j)^2 \ ,
\end{equation}
and consider
\begin{equation}
\begin{split}
\psi_n(x,t,X,T,v,a) &= \psi_n\left(\xi + \frac{a}{\omega^2}\right)  \  e^{i \frac{m a}{\hbar} \left( \frac{a \tau^3}{6}+ \xi  \tau+\frac{\tau ^2 v}{2}\right)}  e^{ i \frac{m v}{\hbar} \left( \frac{\tau  v}{2}+ \xi+X\right)} \  
e^{ - i \frac{\tau  \epsilon }{\hbar }} \ , 
\end{split}
\nonumber 
\end{equation}
where $\tau \equiv  t-T$ and $\psi_n(\xi+ \frac{a}{\omega^2})$ are the eigenfunctions of the shifted sho potential
\begin{equation}
\tilde{V}(\xi) \equiv V(\xi) + m a \xi =  \frac{1}{2} m \omega^2 \left(\xi + \frac{a}{\omega^2} \right)^2 - \frac{m a^2}{2 \omega^2}  \ .
\nonumber
\end{equation}

Upon substitution inside the time--dependent Schr\"odinger equation we find the eigenvalue equation
\begin{equation}
\begin{split}
i \hbar \frac{\partial}{\partial t} \psi_n(x,t,X,T,v,a) &= \left[ -\frac{\hbar^2}{2m} \frac{\partial^2}{\partial x^2} +  V(\xi) + \left(\epsilon -\hbar \omega(n+1/2) + \frac{m a^2}{2\omega^2} \right)  
\right] \psi_n(x,t,X,T,v,a)  \ ,
\end{split}
\nonumber
\end{equation}
which takes a more canonical form after making the identification
\begin{equation}
\epsilon = \tilde{E}_n(a) = \hbar \omega \left(n +\frac{1}{2}\right)- \frac{m a^2}{2 \omega^2} \ . 
\nonumber
\end{equation}

A simple calculation  allows us to express the expectation values of momentum and position in this basis as
\begin{equation}
\begin{split}
\langle p  \rangle &= \langle \hat{p} \rangle_{v=a=0} + m (v+a (t-T)) \\
\langle x  \rangle &= \langle x \rangle_{v=a=0} + \left( X + v (t-T)+ \frac{1}{2} a (t-T)^2-\frac{a}{\omega^2}\right) \ .
\end{split}
\end{equation}

Once again we see that the quantum expectation values contain an intrinsic contribution (i.e. the expectation value in the static potential) and a contribution associated to the motion of the potential.

Also in this case the set $\left\{ \psi_n(x,t,X,T,v,a)  \right\}$ forms a basis associated the Hamiltonian of a potential 
moving with velocity $v$ and acceleration $a$ (the parameter $X$  specifies the position of the potential at a time $T$).

We need to adapt our previous discussion, keeping in mind the qualitative changes introduced by the acceleration.

\begin{itemize}
\item $t< t_1$;
    
\begin{equation}
\begin{split}
\Psi^{(1)}(x,t) &= \sum_l c_l^{(1)} \psi_l(x,t,x_1,t_1,v_1,a_1) \ . \\
\end{split}
\label{eq_t1_accel_1}
\end{equation}

In particular, at $t=t_1$, the expression above becomes
\begin{equation}
\begin{split}
\Psi^{(1)}(x,t_1) &= \sum_l c_l^{(1)} \ \psi_l(x-x_1 + \frac{a_1}{\omega^2}) e^{i \frac{m v_1 (x-x_1)}{\hbar}}  \ , \\
\end{split}
\end{equation}
implying
\begin{equation}
\begin{split}
c_l^{(1)} &= \int_{-\infty}^{\infty} \psi_l^\star(x-x_1 + \frac{a_1}{\omega^2}) \Psi^{(1)}(x,t_1) e^{- i \frac{m v_1 (x-x_1) }{\hbar} } \ dx  \ .
\end{split}
\label{eq_t1_accel_2}
\end{equation}

By using (\ref{eq_t1_accel_2}) inside eq.~(\ref{eq_t1_accel_1}) we obtain 
\begin{equation}
\begin{split}
\Psi^{(1)}(x,t) &= \exp\left\{ i \left(-\frac{a_1 m v_1 \delta t^2 }{\hbar}
+\frac{a_1 m \left(x-x_1\right) \delta t}{\hbar} - \frac{a_1^2 m \delta t^3}{3 \hbar }-\frac{m v_1^2 \delta t}{2 \hbar }
 + \frac{m a_1^2 \delta t}{2 \hbar \omega^2} \right) \right\}  \\
&\cdot \int_{-\infty}^{\infty} K(x-x_1 - v_1 \delta t - \frac{1}{2} a_1 \delta t^2 + \frac{a_1}{\omega^2},t;y-x_1 + \frac{a_1}{\omega^2},t_1)  e^{i \frac{m v_1 (x-y)}{\hbar } }
\Psi^{(1)}(y,t_1) \ dy  \ ,
\end{split}
\label{eq_t1_accel_3} 
\end{equation}
where $\delta t \equiv t-t_1$.

\item $t_1 < t< t_2$;

The set of eigenfunctions of a potential moving with velocity $v$ and acceleration $a$, $\left\{ \psi_n(x,t,X,T,v,a)  \right\}$, constitutes an appropriate basis in which to express the wave function. The natural choice for the parameters $X$ and $T$ are,
as before, $X=x_1$ and $T=t_1$ and $X=x_2$ and $T=t_2$.

In the first case we have
\begin{equation}
\begin{split}
\Psi^{(2)}(x,t) &= \sum_l c_l^{(2)} \psi_l(x,t,x_1,t_1,v_2,a_2)   \ ,
\label{eq_first_acc}
\end{split}
\end{equation}
while in the second case we have
\begin{equation}
\begin{split}
\Psi^{(2)}(x,t) &= \sum_l \bar{c}_l^{(2)} \psi_l(x,t,x_2,t_2,v_2,a_2)   \ .
\label{eq_second_acc}
\end{split}
\end{equation}

At $t=t_1$ we impose the matching condition
\begin{equation}
\Psi^{(1)}(x,t_1) = \Psi^{(2)}(x,t_1) \ ,
\end{equation}
that implies
\begin{equation}
\begin{split}
c_k^{(2)} &= \sum_l \mathcal{Q}_{kl}(v_1,v_2,a_1,a_2;t_1) c_l^{(1)}  \ ,
\end{split}
\nonumber
\end{equation}
where
\begin{equation}
\begin{split}
\mathcal{Q}_{kl}(v_1,v_2,a_1,a_2;t_1) &\equiv \int_{-\infty}^\infty  \psi_k^\star\left(x-x_1 + \frac{a_2}{\omega^2}\right)  \psi_l\left(x-x_1 + \frac{a_1}{\omega^2}\right) e^{i \frac{m (v_1-v_2) (x-x_1)}{\hbar}}  \ dx   \ .
\end{split}
\label{eq_Q_va}
\end{equation}

It is easy to see that these matrix element obey the property (unitarity)
\begin{equation}
\sum_{j=0}^\infty \left| \mathcal{Q}_{k,j}(v_1,v_2,a_1,a_2;t_1)   \right|^2 = 1   \ ,
\nonumber
\end{equation}
which implies the conservation of probability
\begin{equation}
\sum_{k} |c_k^{(2)}|^2 = \sum_{k} |c_k^{(1)}|^2 = 1  \ .
\nonumber
\end{equation}

Equivalently the wave function can be expressed in terms of the quantum propagator as
\begin{equation}
\begin{split}
\Psi^{(2)}(x,t) &=  e^{i \left( - \frac{m a_2^2}{3\hbar} (t-t_1)^3 - \frac{m a_2 v_2}{\hbar} (t-t_1)^2  - \frac{m v_2^2}{2\hbar} (t-t_1) + \frac{m a_2}{\hbar} (x-x_1) (t-t_1) + \frac{m v_2}{\hbar} (x-x_1) + \frac{m a_2^2}{2\omega^2} \right)} \\
&\cdot \int_{-\infty}^\infty K\left(x-x_1- v_2 (t-t_1)- \frac{1}{2} a_2 (t-t_1)^2 +\frac{a_2}{\omega^2}, t; y-x_1+a_2/\omega^2, t_1
\right) \\
&\cdot e^{-i m v_2 (y-x_1)/\hbar} \Psi^{(1)}(y,t_1) dy   \ .
\end{split}
\end{equation}

If we want to be able to iterate this relation for the case of further changes of velocity and acceleration we need to keep in
mind that in deriving it we have used the first basis (that allows the matching at $t=t_1$), while the next change of velocity and acceleration requires specifying the wave function in the second basis.
To do this we need first to calculate $\bar{c}_l^{(2)}$. 

The condition
\begin{equation}
\Psi^{(2)}(x,t) = \sum_l c_l^{(2)} \psi_l(x,t,x_1,t_1,v_2,a_2) = \sum_l \bar{c}_l^{(2)} \psi_l(x,t,x_2,t_2,v_2,a_2) 
\end{equation}
implies
\begin{equation}
\begin{split}
\bar{c}^{(2)}_k &=  \sum_l  c_l^{(2)} e^{\frac{i a_2 m \left(t_1-t_2\right){}^2 v_2}{2 \hbar }} e^{\frac{i a_2^2 m \left(t_1-t_2\right)
        \left(3-\left(t_1-t_2\right){}^2 \omega ^2\right)}{6 \omega ^2\hbar }} e^{\frac{i \left(t_1-t_2\right) \left((2 l+1) \omega  \hbar -m
        v_2^2\right)}{2 \hbar }} \\
& \int_{-\infty}^\infty  e^{-\frac{i a_2 m \left(t_1-t_2\right) y}{\hbar }} \psi_k^\star(y) \psi_l(y) dy  \ .
\end{split}
\end{equation}

Notice that
\begin{equation}
\int_{-\infty}^\infty  e^{-\frac{i a_2 m \left(t_1-t_2\right) y}{\hbar }} \psi_k^\star(y) \psi_l(y) dy = 
\mathcal{Q}_{k,l}(0,-a_2 (t_2-t_1),0,0)
\end{equation}
and therefore
\begin{equation}
\begin{split}
\bar{c}^{(2)}_k 
&=  e^{\frac{i a_2 m \left(t_1-t_2\right){}^2 v_2}{2 \hbar }} e^{\frac{i a_2^2 m \left(t_1-t_2\right) 
\left(3-\left(t_1-t_2\right)^2 \omega ^2\right)}{6 \omega ^2\hbar }}  e^{\frac{i \left(t_1-t_2\right) \left(-m v_2^2\right)}{2 \hbar }} \\
&\cdot \sum_l  c_l^{(2)} e^{\frac{i \left(t_1-t_2\right) \left((2 l+1) \omega  \hbar \right)}{2 \hbar }} \mathcal{Q}_{k,l}(0,-a_2 (t_2-t_1),0,0)  \ .
\label{eq_cbarc}
\end{split}
\end{equation}

The relation between the coefficients $c^{(2)}_k$ and $\bar{c}^{(2)}_k $ is considerably more complicated in presence of an
acceleration: in particular,  a pure state at time $t=t_1^{+}$ will evolve into different states 
at later times $t_1^{+} < t < t_2^{-}$ (in absence of acceleration the state would only pick a phase).

\end{itemize}

\section{Applications}
\label{sec:3}

In this section we will discuss the application of our previous discussion to two physical systems, an infinite square well potential, undergoing multiple sudden changes of velocity, and a quantum harmonic oscillator undergoing multiple sudden changes of velocity and acceleration.

\subsection{Particle in an infinite box under sudden changes of velocity}

We consider a particle in a stationary state of an infinite square well of width $L$ at $t < t_1$; at times $t \leq t_1$ the potential moves with velocity $v_1$, while at time $t=t_1$ it suddenly changes its velocity to $v_2$. 

This sudden change of velocity will induce transitions from the original state to different states of the new potential; the amplitudes for these transitions are obtained from our previous discussion as
\begin{equation}
\begin{split}
\mathcal{Q}_{kl}(v_1,v_2;t_1) &= \left\{
\begin{array}{ccc}
\frac{\pi ^2 k^2 \sin (\delta )}{\pi ^2 \delta  k^2-\delta ^3} & , & k=l \\
i^{k+l}  \frac{16 \pi ^2 \delta  k l \sin \left(\frac{1}{2} \pi  (k+l)-\delta
    \right)}{16 \delta ^4-8 \pi ^2 \delta ^2 \left(k^2+l^2\right)+\pi ^4
    \left(k^2-l^2\right)^2} &, & k \neq l\\
\end{array}
\right.   \ ,
\end{split}
\end{equation}
where we have defined the dimensionless parameter
\begin{equation}
\delta \equiv \frac{L m \left(v_2-v_1\right)}{2 \hbar } \ .
\end{equation}

Let us now consider the following physical situation: at $t \leq t_1$ the particle is in the ground state of the box (not moving); 
at $t=t_1$ the box starts moving with velocity $v$ and finally at time $t=t_2$ the box stops. 
What is the probability that the particle is still in the ground state of the box?

We assume that at time $t_1=0$ the box is centered in the origin ($x_1=0$); $x_2$ is the location of the center of the box at $t=t_2$; clearly
\begin{equation}
x_2 = x_1 + v (t_2-t_1) = v t_2  \ .
\end{equation}

The amplitude of transition for going from the initial ground state to any state at $t>t_2$ is given by
\begin{equation}
c_k^{(3)} = \sum_l \mathcal{Q}_{kl}(v_2,0) \mathcal{Q}_{l,1}(0,v_2) e^{\frac{i m t_2 v_2^2}{2 \hbar }-\frac{i \pi ^2 l^2 t_2 \hbar }{2 L^2 m}}  \ .
\end{equation}

The probability of finding the particle in a state with quantum number $k$ at $t>t_2$ is then 
\begin{equation}
P_k = \left| c_k^{(3)} \right|^2  \ ,
\end{equation}
and it is easy to verify (using the completeness of the basis and the orthonormality of its elements) that 
\begin{equation}
\sum_{k=1}^\infty P_k = 1  \ .
\end{equation}

\begin{figure}
\begin{center}
\includegraphics[width=5.5cm]{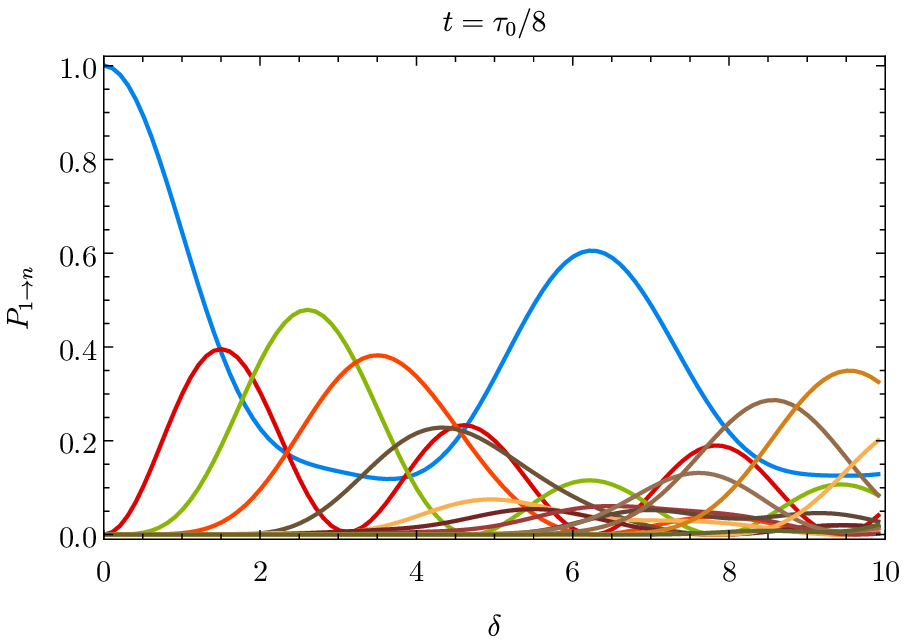} \hspace{.5cm}
\includegraphics[width=5.5cm]{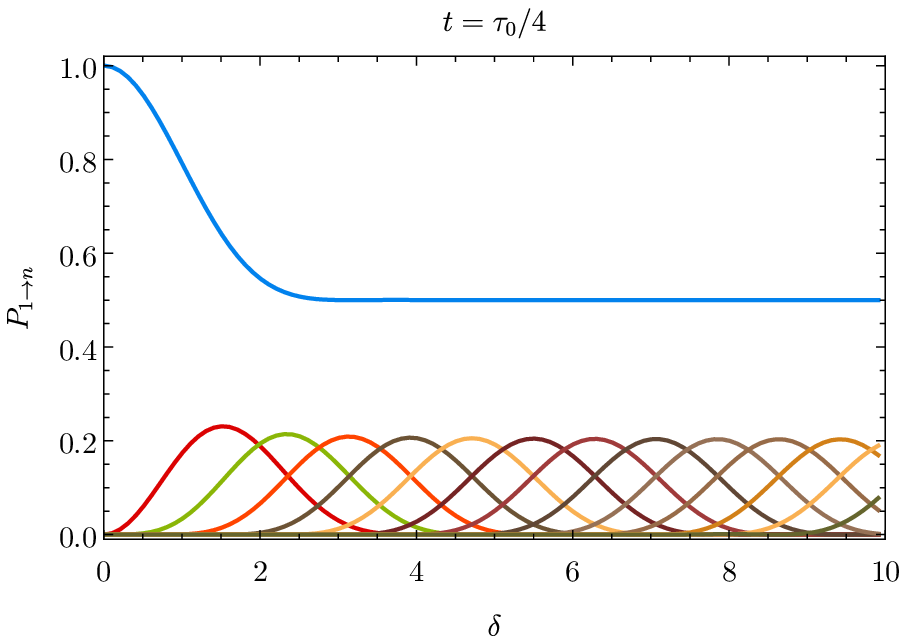} \\
\includegraphics[width=5.5cm]{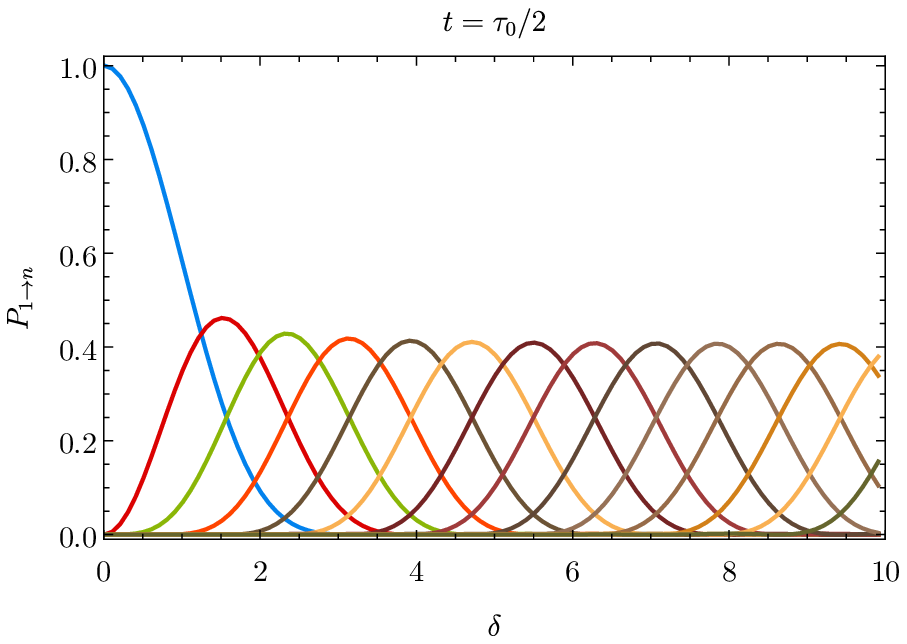} \hspace{.5cm}
\includegraphics[width=5.5cm]{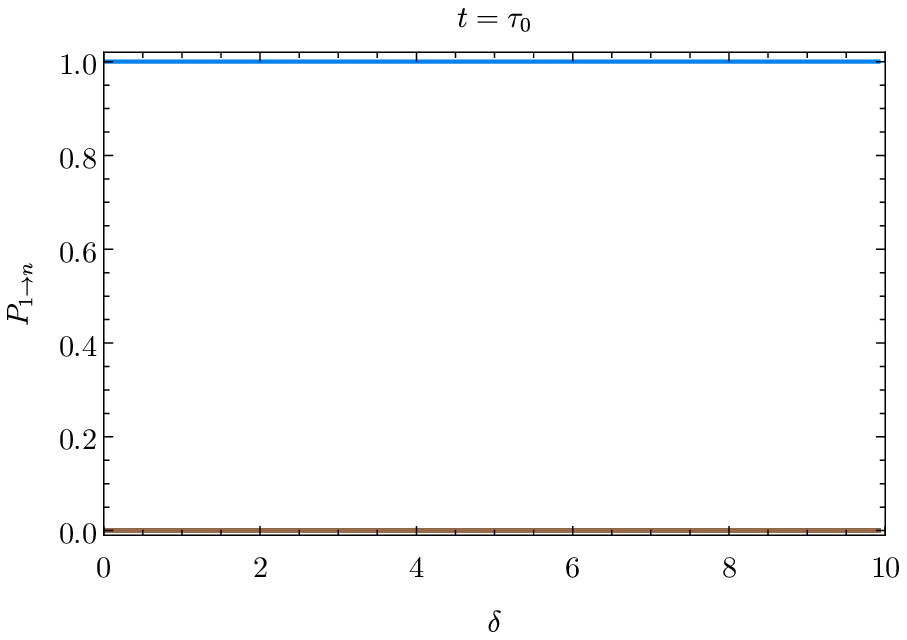} \\
\includegraphics[width=5.5cm]{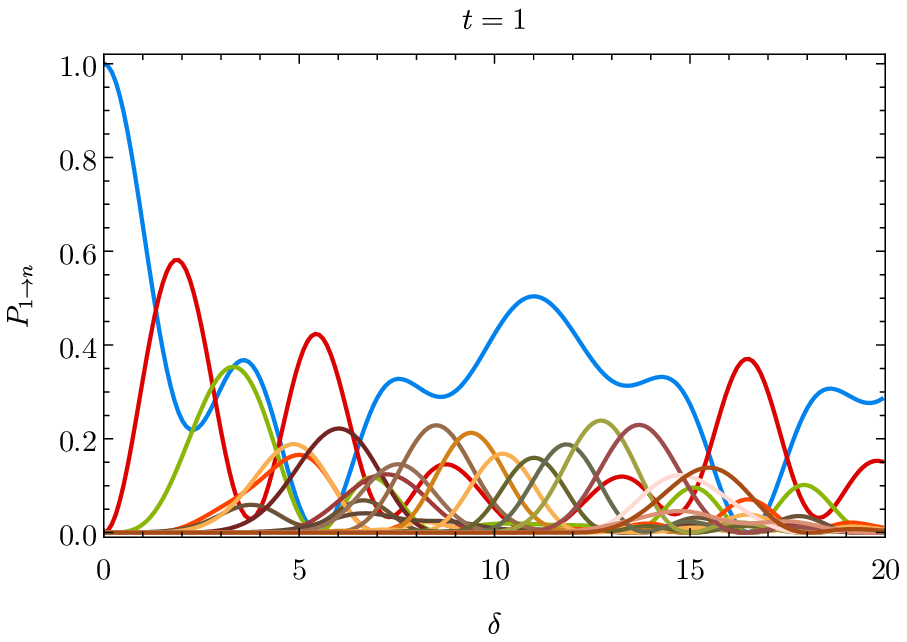} \\
\caption{Probability of transition for a particle initially in the ground state of a static infinite square well that at time $t=0$ starts moving with velocity $v$, and at time $t$ stops ($\tau_0=\frac{4 m L^2}{\pi \hbar} \approx 1.2732$).} 
\label{Fig_SqW}
\end{center}
\end{figure}

In Figs.~\ref{Fig_SqW} we plot the probability of transition for a particle initially in the ground
state of a static infinite square well that starts to move at $t_1=0$ and stops at a time $t$.
In particular at $t=\tau_0 = \frac{4 m L^2}{\pi \hbar}  \approx 1.2732$ the system undergoes a quantum revival and the particle 
is found in the ground state of the potential, regardless of $\delta$~\footnote{The reader interested in the topic of quantum revivals may find refs.~\cite{Styer01,Robinett04} useful.}. 
Notice that $\tau_0/4$ is the time that it takes to a classical particle with velocity $v_0 = \pi \hbar/mL$ to travel a distance $L$.

\begin{figure}
\begin{center}
\includegraphics[width=8cm]{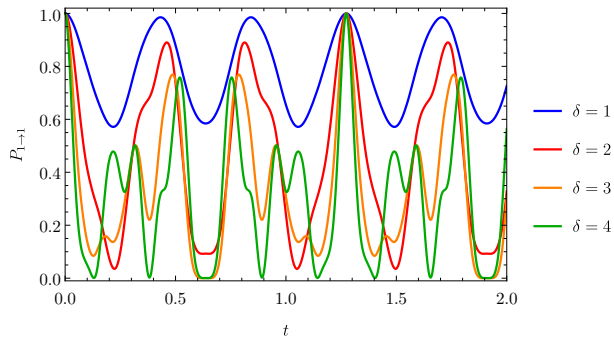} 
\caption{Probability that a particle initially in the ground state of a static infinite square well that at time $t_1=0$ starts moving with velocity $v$ and at time $t_2$ stops, is found in the groud state at times $t>t_2$.} 
\label{Fig_SqW_revival}
\end{center}
\end{figure}

This behavior can be appreciated more clearly in Fig.~\ref{Fig_SqW_revival}, where we have plotted the probability 
$P_{1\rightarrow 1}$ for different values of $\delta$ as a function of the time.

Another interesting behavior is observed at $t=\tau_0/4$, where the particle is found in the ground state with probability $1/2$, for large enough $\delta$.
As we prove next, $1/2$ is just the probability of observing the particle with the final velocity equal to the initial for the corresponding classical system.

Let us now consider the classical system: in this case we have a particle with initial velocity $v_0$ inside a box that is not moving. At time $t_1=0$ the box starts moving with velocity $v>0$ and at a later time $t_2$ the box stops.

We have two possibilities, to start with: $v_0 <0$ and $v_0>0$. 

\begin{itemize}
\item $v_0<0$

In this case, the particle has a relative velocity with respect to the left wall $-v-|v_0|$ and when it hits the wall, its relative velocity gets reversed to $v+|v_0|$. The observer, that sees the potential moving with velocity $v$, in this case sees the particle moving with velocity $2 v+|v_0|$. When the particle reaches the right wall, its velocity gets reversed again and the observer will now see the particle moving with velocity $-|v_0|$.

If the potential is suddenly stopped at a random time the observer will either find that the particle has the same energy it had initially or 
the energy $\frac{1}{2} m (2 v + |v_0|)^2$.

\item $v_0>0$

In this case, the particle has a relative velocity with respect to the left wall $-v+|v_0|$ and when it hits the wall, its relative velocity gets reversed to $v-|v_0|$. The observer, that sees the potential moving with velocity $v$, in this case sees the particle moving with velocity $2 v-|v_0|$. When the particle reaches the right wall, its velocity gets reversed again and the observer will now see the particle moving with velocity $|v_0|$.

If the potential is suddenly stopped at a random time the observer will either find that the particle has the same energy it had initially or the energy $\frac{1}{2} m (2 v -|v_0|)^2$.

\end{itemize}

Since we assume that the potential is stopped at a random time, we need to calculate the statistical average of the 
classical energy, which we call $\mathcal{E}_{cl}$:
\begin{equation}
\mathcal{E}_{cl} = \frac{1}{4} \left[ \frac{\pi ^2 \hbar ^2}{L^2 m}+\frac{1}{2} m \left(2 v-\frac{\pi  \hbar}{L m}\right)^2+\frac{1}{2} m \left(\frac{\pi  \hbar }{L m}+2 v\right)^2 \right]  \ .
\end{equation}

We also define the statistical average of the maximal classical energies $\mathcal{E}^{(+)}_{cl}$:
\begin{equation}
\mathcal{E}^{(+)}_{cl} = \frac{1}{2} \left[ \frac{1}{2} m \left(2 v-\frac{\pi  \hbar}{L m}\right)^2+\frac{1}{2} m \left(\frac{\pi  \hbar }{L m}+2 v\right)^2 \right]  \ .
\end{equation}

\begin{figure}
\begin{center}
\includegraphics[width=5.5cm]{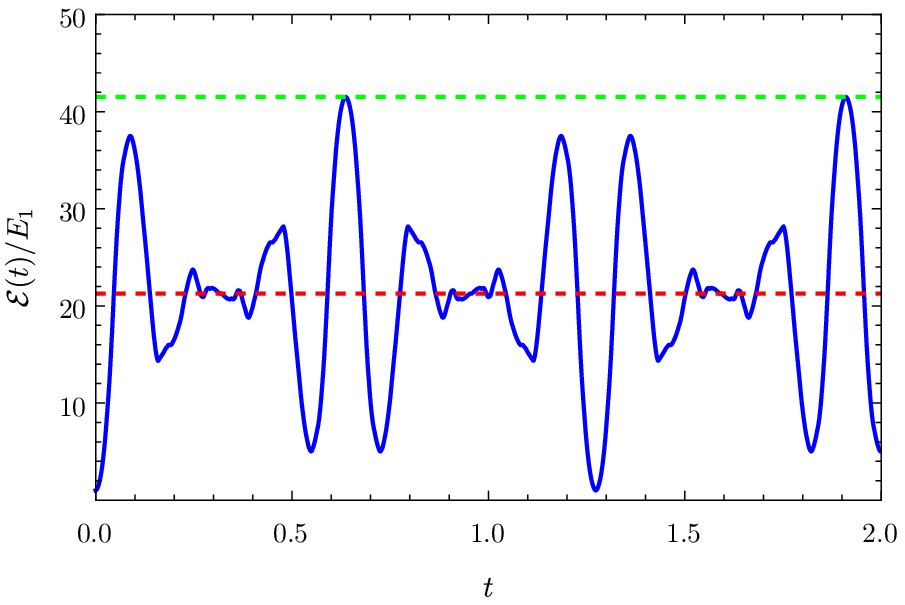} \hspace{0.5cm}
\includegraphics[width=5.5cm]{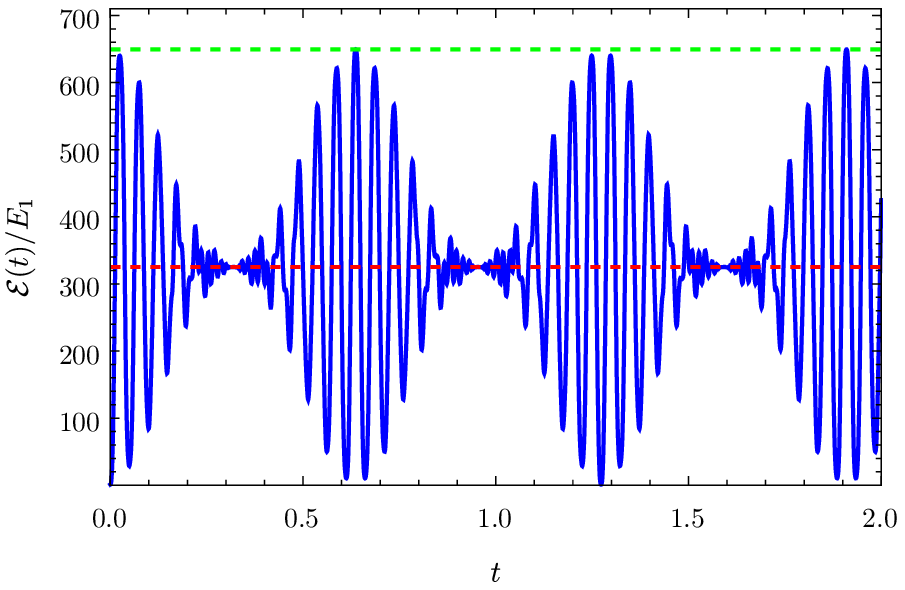} 
\caption{Left plot: Average energy divided by $E_1= \hbar^2 \pi^2/2mL^2$ for  $\delta=m v L/2\hbar = 5$, as a function of time; Right plot: same of left plot for $\delta =20$. The red and green horizontal lines are $\mathcal{E}_{cl}/E_1$ and $\mathcal{E}_{cl}^{(+)}/E_1$.} 
\label{Fig_SqW3}
\end{center}
\end{figure}

In Figs.~\ref{Fig_SqW3} we plot expectation value of the energy of the quantum particle (solid line) and the classical counterpart (red dashed line)
\begin{equation}
\frac{\mathcal{E}_{cl}}{E_1} = 1 + \frac{8 \delta^2}{\pi^2} \ ,
\end{equation}
for $\delta=5$ and $\delta=20$, as a function of the time between the two changes of velocity. The green dashed line corresponds to
\begin{equation}
    \frac{\mathcal{E}^{(+)}_{cl}}{E_1} = 1 + \frac{16 \delta^2}{\pi^2}   \ .
\end{equation}

The quantum expectation value of the energy is seen to oscillate nicely about the statistical average of the classical energy for the two cases in Figs.~\ref{Fig_SqW3}. Additionally, for the case in Fig.~\ref{Fig_SqW} with $t=\tau_0/4$, which corresponds to the case where the classical and quantum expectation values coincide, the quantum probability $P_{1\rightarrow 1} \approx 0.5$, which is also 
the classical probability of observing the particle with the same initial speed.

Consider now a particle in the ground state of a static infinite square well at $t<0$; at time $t_1=0$ the potential (initially centered around $x_1=0$) starts to move with velocity $v>0$, while at time $t=t_2$ (centered around $x_2=x_1+v (t_2-t_1)$) it suddenly reverses its velocity and finally at $t=t_3$ (now centered at $x_3=x_2-v (t_3-t_2)$) it stops.

To make things simpler we will assume $t_3=2 t_2$, so that $x_3=0$. In this case the amplitude for the transition from the ground state of the box to any other state can be calculated as
\begin{equation}
c_l^{(4)} = \exp \left(\frac{i m t v^2}{\hbar }\right) \sum_{k_1,k_2} \mathcal{Q}_{l,k_2}(-v,0) \mathcal{Q}_{k_2,k_1}(v,-v) \mathcal{Q}_{k_1,1}(0,v)  e^{-\frac{i \pi ^2 t \hbar \left({k_1}^2+{k_2}^2\right)}{2 L^2 m}}  \ .
\end{equation}

The same analysis that we have done for the case of two changes of velocities can be carried out also in this case.

In particular the previous discussion for the classical system can be extended to the present case.
In this case we find that at the end (after a time $2t$) an external observer can detect the speeds: $v$ (2)  , $2 v + v_0$ (2),
$2 v-v_0$ (2), $4 v-v_0$ (1) and $4 v+v_0$(1). In parenthesis are the degenerations of these possibilities.

We then find that
\begin{equation}
\begin{split}
\frac{\mathcal{E}_{cl}}{E_1} &= 1+\frac{24 \delta ^2}{\pi ^2} \\
\frac{\mathcal{E}_{cl}^{(+)}}{E_1} &= 1+\frac{64 \delta ^2}{\pi ^2}  \ .
\end{split}
\end{equation}

\begin{figure}
\begin{center}
\includegraphics[width=5.5cm]{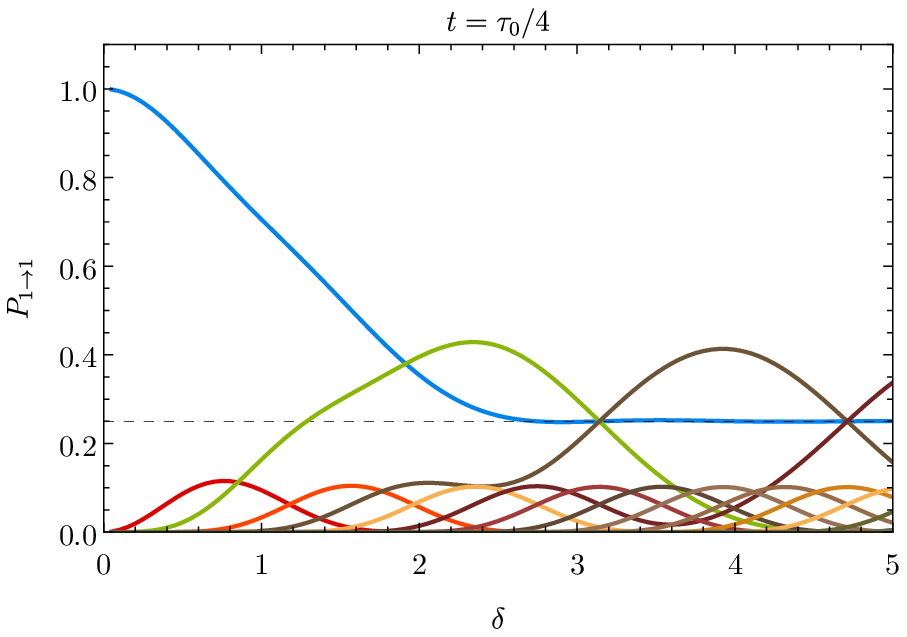} \hspace{.5cm}
\includegraphics[width=5.5cm]{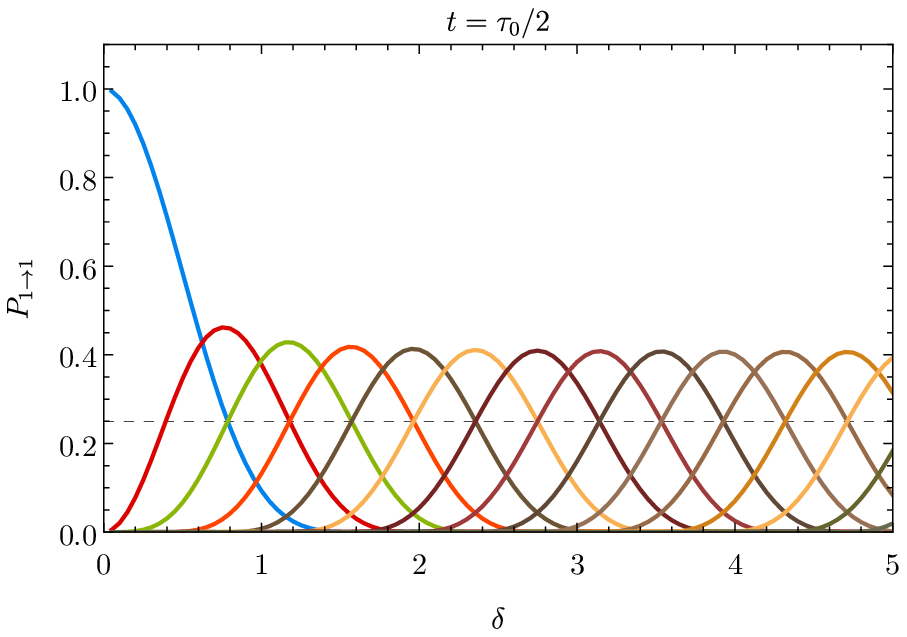} \\
\includegraphics[width=5.5cm]{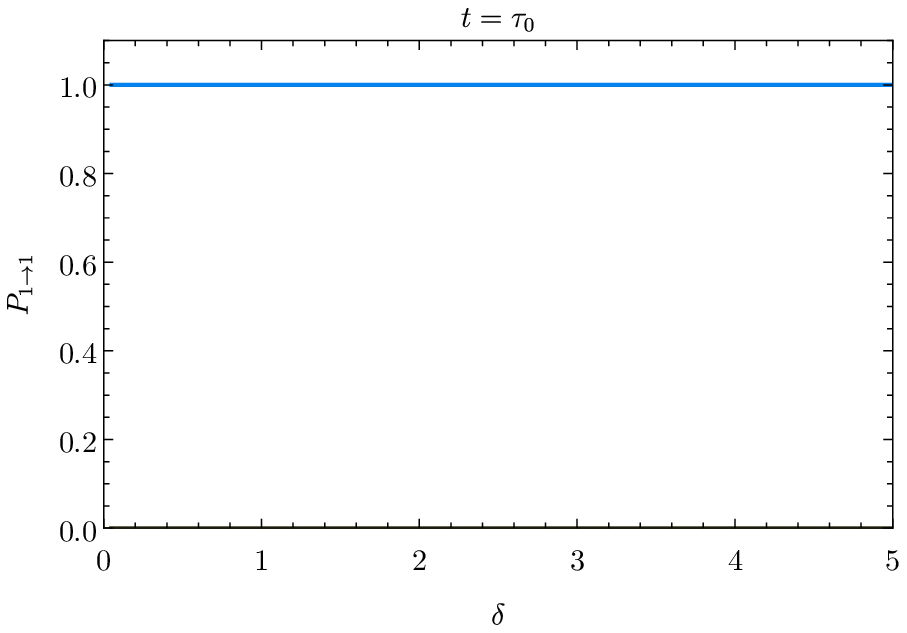} \hspace{.5cm}
\includegraphics[width=5.5cm]{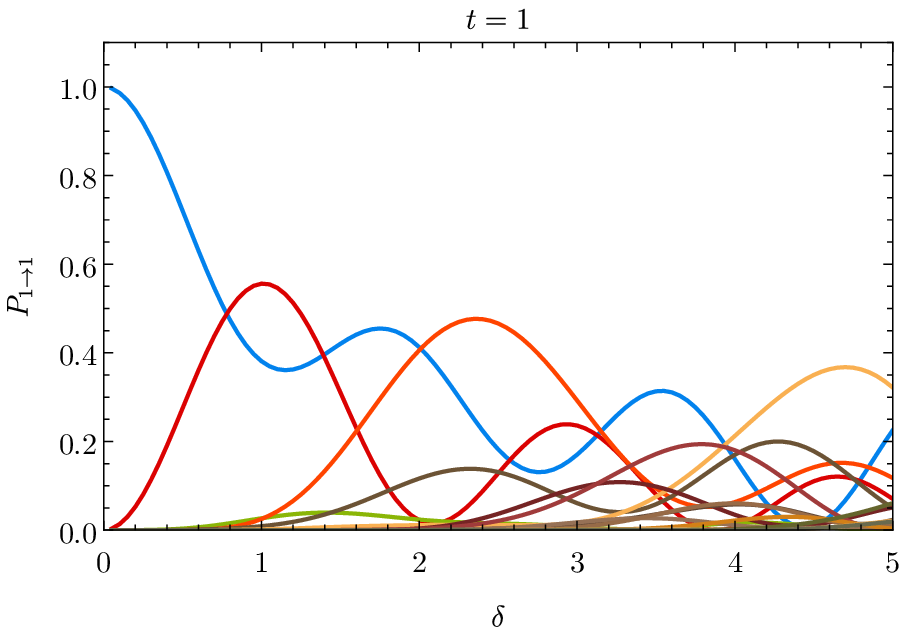} \\
\caption{Probability of transition for a particle initially in the ground state of a static infinite square well 
that at time $t=0$ starts moving with velocity $v$, and at time $t$ inverts its velocity and at time $2t$ stops. ($\tau_0= 4 mL^2/\hbar \pi$).} 
\label{Fig_SqW_3changes}
\end{center}
\end{figure}

In Fig.~\ref{Fig_SqW_3changes} we plot the probability of transition from the ground state to any other state
under a triple change of velocity. Once again we see that  at $t=\tau_0$ the system undergoes a quantum revival.

\begin{figure}
\begin{center}
\includegraphics[width=8cm]{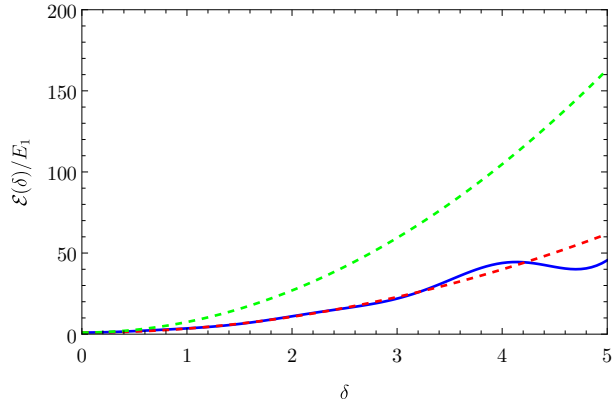} 
\caption{Expectation value of the energy (divided by $E_1$) as a function $\delta$ for $t= \tau_0/4$. 
The red and green lines are $\mathcal{E}_{cl}/E_1$ and $\mathcal{E}_{cl}^{(+)}/E_1$ respectively.} 
\label{Fig_SqW_3changes_energy}
\end{center}
\end{figure}

At $t=\tau_0/4$ the system is the most classical, as one can see from Fig.~\ref{Fig_SqW_3changes_energy}: this is also appreciated in 
the first plot of Fig.~\ref{Fig_SqW_3changes}, where for $\delta$ sufficiently large the probability of staying in the ground state tends
to $1/4$ (horizontal dashed line). Actually, $1/4$ is precisely the classical probability of observing the particle with the same speed, if we do not know its original position (a similar situation occurred for two changes of speed, where in that case the probability is $1/2$). 

In the general case of $n$ changes of velocity, we may expect that the classical probability of finding the particle still in the same state to be $1/2^{n-1}$.

\begin{figure}
\begin{center}
\includegraphics[width=8cm]{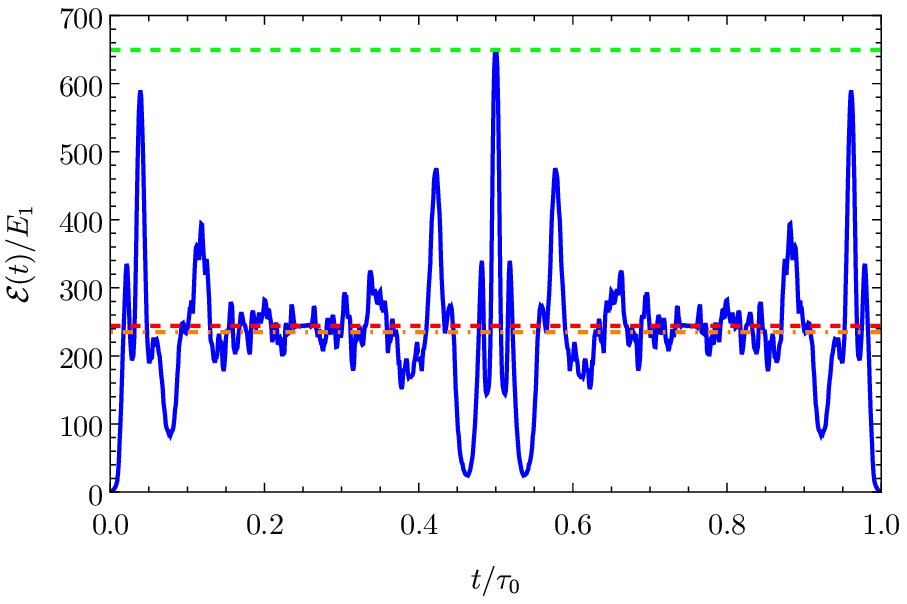} 
\caption{Average energy divided by $E_1= \hbar^2 \pi^2/2mL^2$ for  $\delta=m v L/2\hbar = 10$, as a function of time, summing over $50$ states. 
The horizontal red and green lines are $\mathcal{E}_{cl}/E_1$ and $\mathcal{E}_{cl}^{(+)}/E_1$. 
The horizontal orange line (dotdashed) is the time average of the quantum expectation value over the time $\tau_0$: $\frac{1}{\tau_0} \int_0^{\tau_0} \frac{\mathcal{E}_{cl}^{(+)}}{E_1} dt$. } 
\label{Fig_SqW6}
\end{center}
\end{figure}

As we can see from Fig.~\ref{Fig_SqW6}, the quantum expectation value of the energy oscillates about 
the classical value $\mathcal{E}_{cl}$ (red dashed line) and its maximum value is very close to $\mathcal{E}_{cl}^{(+)}$ (green dashed line).

\subsection{Simple harmonic oscillator}

Let us imagine that at time $t<t_1$ the particle is in a stationary state of the sho potential with zero velocity and acceleration ($v_1=0$ and $a_1=0$): in particular at $t=t_1$ we have (we use $x_1=0$)
\begin{equation} 
\Psi(x,t_1) = \psi_{n_0}(x) 
\end{equation}
and
\begin{equation}
c_{l}^{(1)} = \delta_{l,n_0} \ .
\end{equation}

For $t>t_1$ the potential starts to move with velocity $v_2$ and acceleration $a_2$: the amplitude for the transition to a given state of the moving potential are given by
\begin{equation}
c_{k}^{(2)} = \sum_{l} \mathcal{Q}_{kl}(v_1,v_2,a_1,a_2;t_1) c_{l}^{(1)} \ ,
\end{equation}
where the matrix elements $\mathcal{Q}_{kl}(v_1,v_2,a_1,a_2;t_1) c_{l}^{(1)}$ have been defined in eq.~(\ref{eq_Q_va}).
For the special case of simple harmonic oscillator these matrix elements can be expressed as
\begin{equation}
\begin{split}
\mathcal{Q}_{kl}(v_1,v_2,a_1,a_2;t_1) &= \frac{ e^{\frac{i \kappa  \rho }{2}-\frac{\lambda ^2}{4}}}{\sqrt{\pi} \sqrt{2^k k!} \sqrt{2^l l!}}   \nonumber \\
&\cdot \int_{-\infty}^\infty  H_k\left(x+\frac{\lambda }{2}\right) \  H_l\left(x-\frac{\lambda }{2}\right) \ e^{-x^2-i \kappa  x} dx    \ ,
\end{split}
\end{equation}
where we have introduced the dimensionless variables
\begin{equation}
\begin{split}
\kappa &\equiv \sqrt{\frac{m}{\hbar \omega}} (v_2-v_1) \\
\lambda &\equiv \sqrt{\frac{m}{\hbar \omega^3}} (a_2-a_1) \\
\rho &\equiv \sqrt{\frac{m}{\hbar \omega^3}} (a_2+a_1)  \ . \\
\end{split}
\end{equation}

We have found out that $\mathcal{Q}_{k,j}(v_1,v_2,a_1,a_2;t_1) $ takes the general form (for a derivation see the appendix)
\begin{equation}
\begin{split}
\mathcal{Q}_{k,j}(v_1,v_2,a_1,a_2;t_1)  &= 
\left\{
\begin{array}{ccc}
e^{-\left(\kappa^2+\lambda^2-2 i \kappa \rho\right)/4} L_j(\frac{\kappa^2+\lambda^2}{2}) & , & k =j \\
- \left(\frac{i (\kappa-i (-1)^{\vartheta(k-j)} \lambda) }{\sqrt{2}}\right)^{|k-j|}  \ \frac{\sqrt{\frac{\min (j,k)!}{\max (j,k)!}}}{\Gamma(|k-j|)} \ e^{-\left(\kappa^2+\lambda^2-2 i \kappa \rho\right)/4}  & & \\
\cdot \sum _{l=0}^{\min (j,k)} \left(\prod _{p=1}^{| k-j| -1} (l-\min (k,j)-p)\right)
L_l(\frac{\kappa^2+\lambda^2}{2}) & , & |k-j|>0 \\
\end{array}\right.       \ ,
\end{split}
\end{equation}
where $L_j(x)$ is the Laguerre polynomial of order $j$, $\Gamma(x)$ is the Euler gamma function and $\vartheta(x)$ is the Heaviside theta function. Also observe that $\left| \mathcal{Q}_{k,j}(v_1,v_2,a_1,a_2;t_1)   \right|^2$ is independent of $\rho$ and that $\mathcal{Q}_{k,j}(v_1,v_2,a_1,a_2;t_1)  =  \mathcal{Q}_{j,k}(v_1,v_2,a_1,a_2;t_1)$.

This formula can be used inside eq.~(\ref{eq_cbarc}) to calculate the probability of transition for a quantum particle in an eigenstate 
of a uniformly accelerating (with acceleration $a$) SHO between the times $t_1$ and $t_2$ as~\footnote{Notice that no change of acceleration is involved.} 
\begin{equation}
P_{i \rightarrow f} = \left| \mathcal{Q}_{f,i}(0,-a (t_2-t_1),0,0 )\right|^2 \ .
\end{equation} 

By using the explicit expression for $\mathcal{Q}$ we find
\begin{equation}
P_{i \rightarrow f} = \frac{\min (i,f)!}{\max (i,f)!} \gamma^{|f-i|}  e^{-\gamma} \left[ L_{\min (i,f)}^{|i-f|}(\gamma)\right]^2 \ ,
\label{eq_prob_lud}
\end{equation} 
with $\gamma= \frac{m a^2 t^2}{2\hbar \omega}$. The equation above formally agrees with eq.~(3) of \cite{Dodonov21}, which was originally derived by Ludwig~\cite{Ludwig51}, although with a different $\gamma$. The general expression  for $\gamma$ found by Ludwig, given in eq.(4) of \cite{Dodonov21}, has been deemed incorrect by Dodonov, who has proposed an alternative expression (see eq.(12) of \cite{Dodonov21}):  for the special case of a SHO,  $\gamma_{\rm Dodonov} = \frac{2 m a^2}{\hbar \omega^3} \sin^2(\omega t/2)$, which reduces to our expression for $\omega t \ll 1$, but differs otherwise (we do not find periodic excitations due to constant acceleration). Interestingly, Schwinger finds the transition probability for the electromagnetic field perturbed by a time dependent current to be of the form (\ref{eq_prob_lud}) (see eq.(39) of Ref.~\cite{Schwinger53}), independently of Ludwig.

We now consider the case of a single change of acceleration or velocity; if the particle is initially in the ground state  of the SHO,  
under a single change of velocity and acceleration the wave function will have coefficients 
\begin{equation}
c_{l}^{(2)} = \frac{1}{\sqrt{l!}} e^{-\frac{1}{4} \left(\kappa^2-2 i \kappa  \rho +\lambda ^2\right)} 
\left( \frac{\lambda -i \kappa }{\sqrt{2}} \right)^l \ ,
\label{eq_cl2}
\end{equation}
and  the probability of finding an excited state with quantum number $l$ will be
\begin{equation}
P_{0 \rightarrow l} = \left| c_{l}^{(2)}\right|^2 = e^{-(\kappa^2+\lambda^2)/2} \frac{1}{l!} \left( \frac{\kappa^2+\lambda^2}{2}\right)^l \ .
\end{equation}

It is important to observe that  the state with coefficients (\ref{eq_cl2}) is a {\sl coherent}.
Indeed, using the coefficients of eq.(\ref{eq_cl2}) and the formulas (\ref{eq_Delta_x}), we find 
that wave packet propagates without spreading
\begin{equation}
\Delta x   =   \sqrt{\frac{\hbar}{2 m \omega}} \  ,
\end{equation}
and it is peaked around the expectation value
\begin{equation}
\begin{split}
\langle x \rangle &= \left( x_1 + v_2 (t-t_1) + \frac{a_2}{2} (t-t_1^2)\right) +\sqrt{\frac{\hbar }{m \omega }} (\lambda \cos (\omega  (t-t_1))-\kappa  \sin(\omega  (t-t_1)))  \ .
\label{eq_xave}
\end{split}
\end{equation}

The first term corresponds to the position of the minimum of the moving potential, whereas the second term corresponds to 
an oscillatory movement induced by the change of velocity and acceleration.

\begin{figure}
\begin{center}
\includegraphics[width=7cm]{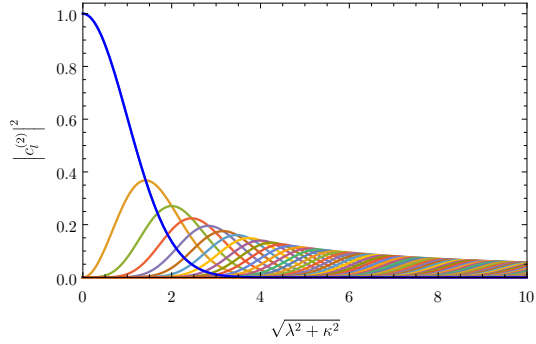} 
\caption{Probability of transition $P_{0 \rightarrow l}$ for a particle initially in the ground state of a static SHO as a function of $\sqrt{\lambda^2+\kappa^2}$.}
\label{Fig_one_change}
\end{center}
\end{figure}

It is easy to see that the maximum of $|c_l^{(2)}|^2$ occurs at $\sqrt{\kappa^2+\lambda^2} = \sqrt{2 (l-1)}$, which corresponds to the condition
\begin{equation}
\frac{1}{2} m v^2 + \frac{m a^2}{2 \omega^2}  = \hbar \omega (l-1) \ .
\end{equation}

A similar experiment can be carried out for the case where the particle is initially in the first excited state of the SHO; in this case we 
find
\begin{equation}
c_{l}^{(2)} = \frac{2^{-\frac{l}{2}-\frac{1}{2}}}{\sqrt{l!}} 
e^{\frac{-\kappa ^2+2 i \kappa  \rho -\lambda ^2}{4} } (\lambda -i
   \kappa )^{l-1} \left(-\kappa ^2-\lambda ^2+2 l\right)
\label{eq_cl2b}
\end{equation}
and
\begin{equation}
P_{1 \rightarrow l} =  e^{-(\kappa^2+\lambda^2)/2}
\frac{1}{l!}    \left(\frac{\kappa ^2+\lambda ^2}{2} \right)^{l-1} 
\left(\frac{\kappa ^2+\lambda ^2}{2}- l\right)^2 \ .
\end{equation}
Notice that the value $\eta= \sqrt{2l}$ corresponds to a zero of the probability, while the probability now displays two peaks.
Using the coefficients above we also find in this case that the wave packet propagates without spreading, with a constant width
\begin{equation}
\Delta x   =   \sqrt{\frac{3 \hbar}{2 m \omega}} 
\end{equation}
and it is peaked around the expectation value (\ref{eq_xave}).

Similarly for the particle initially in the second excited state we have
\begin{equation}
\begin{split}
c_{l}^{(2)} &= -\frac{2^{\frac{1}{2} (-l-3)} e^{\frac{1}{4} \left(-\kappa ^2+2 i
   \kappa  \rho -\lambda ^2\right)} (\lambda -i \kappa )^l
   \left(\left(\kappa ^2+\lambda ^2\right)^2+4 l^2-4 l \left(\kappa
   ^2+\lambda ^2+1\right)\right)}{\sqrt{l!} (\kappa +i \lambda )^2} 
\end{split}
\end{equation}
and
\begin{equation}
\begin{split}
P_{2 \rightarrow l} &= \frac{2^{-l-3}}{l!}  e^{-\left(\kappa ^2+\lambda^2\right)/2}
   \left(\kappa ^2+\lambda ^2\right)^{l-2} \left(\left(\kappa
   ^2+\lambda ^2\right)^2+4 l^2-4 l \left(\kappa ^2+\lambda
   ^2+1\right)\right)^2 \ .
\end{split}
\end{equation}
Notice that the probability has now three peaks for $l \geq 2$.

Also in this case the wave packet doesn't spread
\begin{equation}
\Delta x   =   \sqrt{\frac{5 \hbar}{2 m \omega}}  \ ,
\end{equation}
and it is peaked around the expectation value (\ref{eq_xave}).

We now consider the case of a particle in the ground state of the simple harmonic oscillator a $t < t_1=0$. 
The potential is not moving ($v_1=0$ and $a_1=0$). For  $t_1 \leq t \leq t_2$ the potential moves with velocity $v_2$ and acceleration $a_2$ and at $t=t_2$ it suddenly stops ($v_3=0$ and $a_3=0$).

At $t > t_3$ the wave function of the particle  will be $\Psi = \sum_{j=0}^\infty c_j^{(3)} \psi_j$, where 
\begin{equation}
\begin{split}
c_j^{(3)} &= \sum_{k=0}^\infty \sum_{l=0}^\infty \exp \left(\frac{i a_2^2 m \tau ^3}{6 \hbar}-\frac{i a_2^2 m \tau }{2 \omega ^2 \hbar}+\frac{i a_2 m \tau ^2 v_2}{2 \hbar }-
i (l+1/2) \tau \omega +\frac{i m \tau  v_2^2}{2 \hbar }\right) \\
&\cdot \mathcal{Q}_{jk}(v_2,0,a_2,0) \mathcal{Q}_{kl}(0,-a_2 (t_2-t_1),0,0) \mathcal{Q}_{l0}(0,v_2,0,a_2) \ ,
\end{split}
\end{equation}
and $\tau \equiv t_2-t_1$. Notice that if the potential does not accelerate ($a_2=0$)
$ \mathcal{Q}_{kl}(0,0,0,0)= \delta_{kl}$ and the coefficient takes a much simpler form. 

\begin{figure}
\begin{center}
\includegraphics[width=9cm]{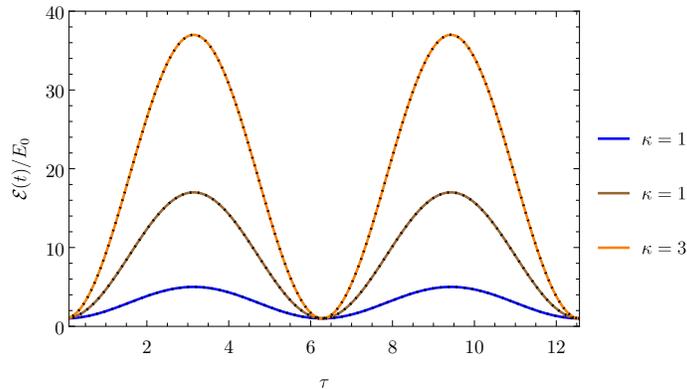} 
\caption{$\mathcal{E}(t)/E_0$ for a quantum oscillator initially in the ground state for $t\leq t_1$ and moving with 
velocity $\kappa = 1$ (blue line), $\kappa = 2$ (brown line) and $\kappa = 3$ (orange line) for $t_1 \leq t \leq t_2=t_1+\tau$  and suddenly stopping at $t_2$. $E_0 = \hbar \omega/2$.
The dotted lines are the classical value calculated in eq.~(\ref{eq_sho_cl}). }
\label{Fig_eave_sho_v}
\end{center}
\end{figure}

We will first consider this case,  and calculate the expectation value of the energy at times 
$t > t_3$ as a function of $\tau$. The expectation value of the energy (divided by $E_0$) is displayed in Fig.~\ref{Fig_eave_sho_v}, for three different
values of $\kappa$ (solid lines); the dotted lines that nicely overlap with this curves represent the classical expectation value of the energy 
of eq.(\ref{eq_sho_cl}) for these values of $\kappa$. Observe that in this case the system undergoes periodic excitations (and actually the classical energy is actually a periodic function).

\begin{figure}
\begin{center}
\includegraphics[width=5.5cm]{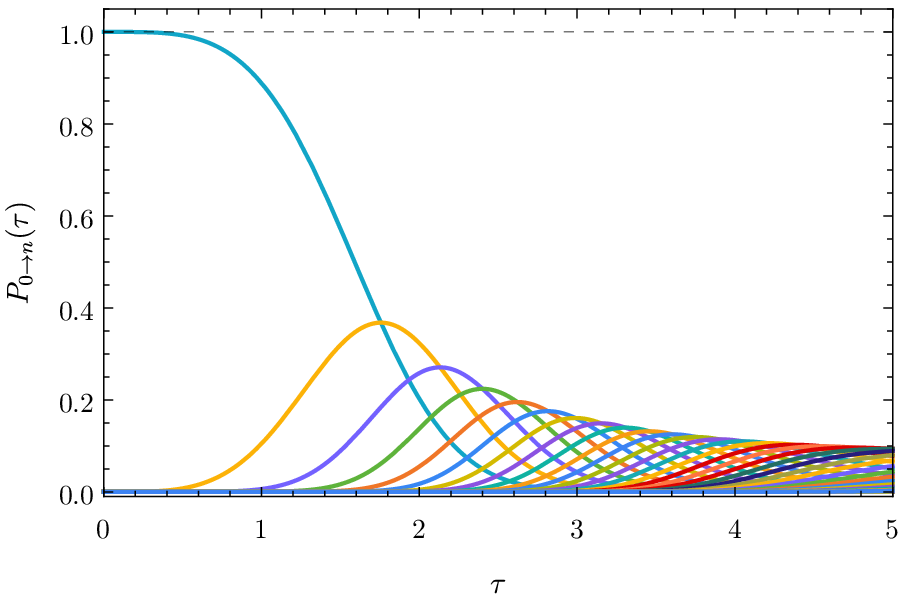} \hspace{0.5cm}
\includegraphics[width=5.5cm]{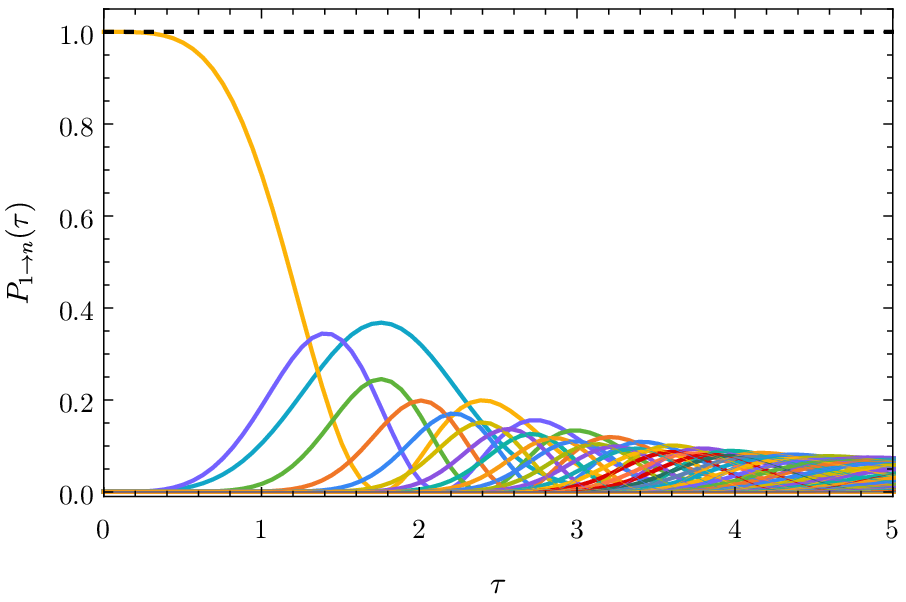} \caption{Probability of transition from the ground state to any other state due to two sudden changes of acceleration ($a_1=a_3=0$, $a_2 = \frac{\omega ^{3/2} \sqrt{\hbar }}{\sqrt{m}}$) as a function  of the time $\tau$ between the two changes. The horizontal dashed line is the numerical sum of the probabilities up to the $40$ state. The left (right) plot corresponds to having initially the particle in the $n=0$ ($n=1$) eigenstate of the oscillator. }
\label{Fig_SHO_a_2changes_p}
\end{center}
\end{figure}

\begin{figure}
\begin{center}
\includegraphics[width=5cm]{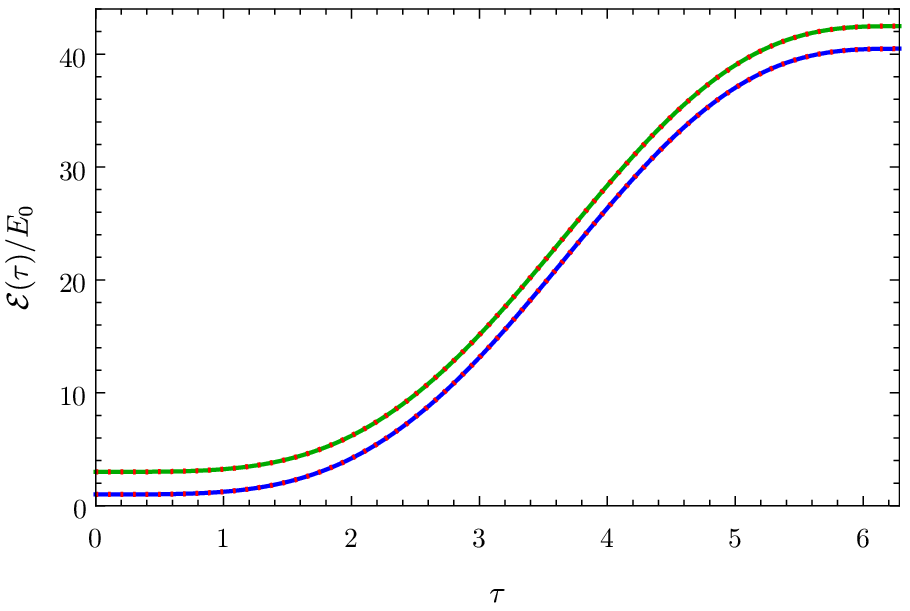} \hspace{1cm} 
\includegraphics[width=5cm]{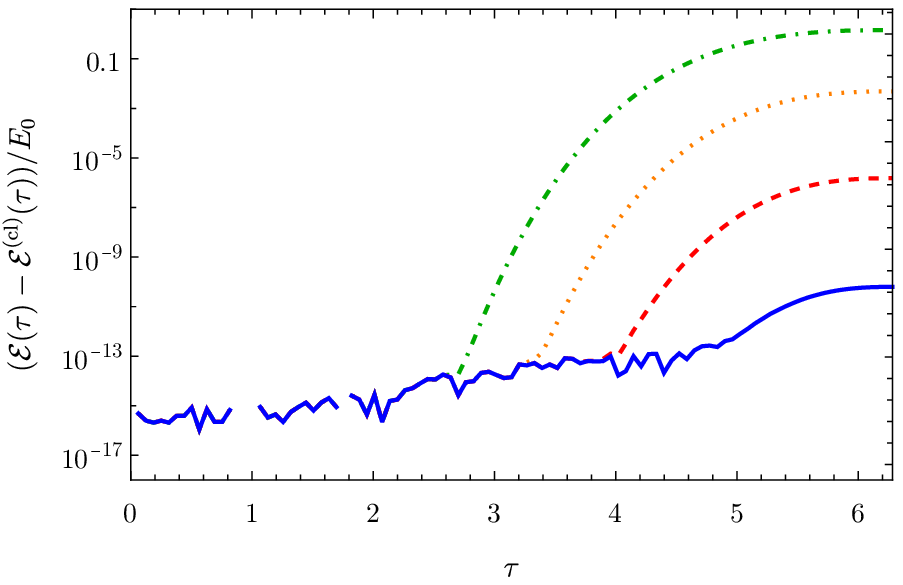} 
\caption{Left plot: $\mathcal{E}(t)/E_0$ for a quantum oscillator initially in the ground state  (blue line) or in the first excited state (green line). The potential undergoes a change in acceleration at $t=t_1$,  going from $a_1=0$ to $a_2 = \frac{\omega ^{3/2} \sqrt{\hbar }}{\sqrt{m}}$; at $t=t_2$ the acceleration suddenly stops, $a_3=0$. The dotted red lines are the  classical value obtained from eq.~(\ref{eq_sho_cl}); Right plot: $(\mathcal{E}(t)-\mathcal{E}^{(cl)}(t))/E_0$ for the first case  plotted in Fig.~\ref{Fig_SHO_a_2changes_e}, using $30$, $40$, $50$ and $60$ (in order of decreasing values).}
\label{Fig_SHO_a_2changes_e}
\end{center}
\end{figure}

This analysis has also been extended to the case of a SHO which suddenly accelerates between $t=0$ and $t=\tau$, while being at rest otherwise.
The left (right) Fig.~\ref{Fig_SHO_a_2changes_p} shows the probability of transition from the ground (first excited) state to any other state of the SHO  due to two sudden changes of acceleration ($a_1=a_3=0$, $a_2 = \frac{\omega^{3/2} \sqrt{\hbar }}{\sqrt{m}}$) as a function  of the time $\tau$ between the two changes. The corresponding expectation values of the energy for the two cases are displayed in the left plot of Fig.~\ref{Fig_SHO_a_2changes_e}, where the solid lines (blue and green) represent the quantum result and the dashed lines correspond to the classical expression obtain in the Appendix.
The quantum result is essentially indistinguishable from the classical result, something that is better appreciated  by looking at the right plot of  Fig.~\ref{Fig_SHO_a_2changes_e}, where we plot 
$(\mathcal{E}(t)-\mathcal{E}^{(cl)}(t))/E_0$ for the case of a particle initially in the ground state of a static SHO at $t\leq 0$, if the potential suddenly uniformly accelerates 
for $0<t< \tau$. The different curves correspond to using $30$, $40$, $50$ and $60$ states in the sums (in decreasing order of the plotted quantity).
 
\section{Conclusions}
\label{sec:4}

In this paper we have considered the problem of a quantum particle in a potential that can suddenly change velocity and acceleration multiple times.
It is remarkable, in our view, that this problem had not been considered earlier, also taking into account its relevance in describing possible experimental situations (for example the translation of a quantum particle trapped in a potential generated by an optical tweezer). 
The paper by Dodonov, \cite{Dodonov21}, which has a rich and interesting list of references, analyses the related problem of a potential (simple harmonic oscillator) moving under a general law (no sudden changes). Our results for constant acceleration (no change of acceleration) reproduce the functional form for the probability of transition originally derived by Ludwig long time ago~\cite{Ludwig51}, although with a different expression 
for the parameter $\gamma$. Interestingly, one of the main results obtained by Dodonov in \cite{Dodonov21} is to provide a different expression 
for $\gamma$: our expression and Dodonov's agree for $\omega t \ll 1$, but lead to different physical predictions otherwise (in our case there is no periodic excitation).

The analysis for two or more changes of velocity and acceleration presented in this paper is based on a rigorous use of spectral decomposition in the basis of the moving potential and it can be applied to general potentials in one or more dimensions, including potentials with a mixed spectrum (discrete and continuum).  
For the simple harmonic oscillator under sudden changes of velocity and acceleration we have found that the expectation value of the energy for the wave packet generated  in this process is very close to the classical expectation value (actually identical within numerical error), an outcome that we interpret as a manifestation of the coherence of the wavepacket.

\section*{Acknowledgements}
The research of P.A. was supported by Sistema Nacional de Investigadores (M\'exico). 
P.A. acknowledges fruitful discussions with dr A. Aranda, dr M. Rodriguez-Vega and dr M. Tejeda-Yeomans.

\appendix

\appendix

\section{Amplitude of transitions}

The purpose of this appendix is the calculation of integrals of the form
\begin{equation}
Q_{mn}(\alpha ,\beta )=\int_{-\infty }^{\infty }\psi _{m}(x+\alpha )^{*}\psi
_{n}(x)\,e^{\beta x}\,dx,  \label{eq:Q_(mn)_inte}
\end{equation}
where $\psi _{j}(x)$ is an eigenfunction of the dimensionless Hamiltonian
operator $H_{HO}=p^{2}/2+x^{2}/2$, $[x,p]=i$.

If we write $\psi _{m}(x+\alpha )=e^{\alpha D}\psi _{m}(x)$, where $D=d/dx$,
then the integral (\ref{eq:Q_(mn)_inte}) can be rewritten
\begin{equation}
Q_{mn}(\alpha ,\beta )=\left\langle \psi _{m}\right| e^{-\alpha D}e^{\beta
x}\left| \psi _{n}\right\rangle ,  \label{eq:Q_(mn)_bracket_op}
\end{equation}
if $\alpha $ is real.

We can rewrite the product of exponential operators in a convenient way by
means of the Baker-Campbell-Hausdorff (BCH) formula. If $A$ and $B$ are two
linear operators that commute with $[A,B]$ then the BCH formula states that
\begin{equation}
e^{A+B}=e^{A}e^{B}e^{[A,B]/2}.  \label{eq:BCH}
\end{equation}
In the present case we resort to the expressions $x=\left( a+a^{\dagger
}\right) /\sqrt{2}$ and $D=\left( a-a^{\dagger }\right) /\sqrt{2}$ where $a$
and $a^{\dagger }$ are the annihilation and creation operators,
respectively, that satisfy $[a,a^{\dagger }]=1$. Then we have
\begin{eqnarray}
e^{-\alpha D}e^{\beta x} &=&e^{-\alpha D+\beta x}e^{-\alpha \beta /2}=\exp
\left( \frac{\beta -\alpha }{\sqrt{2}}a+\frac{\alpha +\beta }{\sqrt{2}}%
a^{\dagger }\right) \exp \left( -\frac{\alpha \beta }{2}\right)   \nonumber
\\
&=&\exp \left( \frac{\alpha +\beta }{\sqrt{2}}a^{\dagger }\right) \exp
\left( \frac{\beta -\alpha }{\sqrt{2}}a\right) \exp \left( \frac{\beta^{2}-\alpha ^{2}}{4}-\frac{\alpha \beta }{2}\right) .
\label{eq:disentangling}
\end{eqnarray}

Taking into account that $a\psi _{n}=\sqrt{n}\psi _{n-1}$ we obtain
\begin{equation}
e^{\xi a/\sqrt{2}}\psi _{n}=\sum_{j=0}^{n}\frac{\xi ^{j}}{j!2^{j/2}}%
R_{nj}\psi _{n-j},\;R_{nj}=\sqrt{n(n-1)\ldots (n-j+1)}.  \label{eq:exp(a)_psi}
\end{equation}
Therefore, it follows from
\begin{equation}
Q_{mn}=\left\langle \exp \left( \frac{\alpha +\beta ^{*}}{\sqrt{2}}a\right)
\psi _{m}\right| \left. \exp \left( \frac{\beta -\alpha }{\sqrt{2}}a\right)
\psi _{n}\right\rangle \exp \left( \frac{\beta ^{2}-\alpha ^{2}}{4}
-\frac{\alpha \beta }{2}\right)   \label{eq:Q_(mn)_disentangling}
\end{equation}
that
\begin{eqnarray}
Q_{mn} &=&\exp \left( \frac{\beta ^{2}-\alpha ^{2}}{4}-\frac{\alpha \beta }{2}\right)   \nonumber \\
&&\times \sum_{j=0}^{m} \sum_{k=0}^{n} \frac{\left( \alpha +\beta \right)^{j} \left( \beta -\alpha \right)^{k}}{2^{(j+k)/2}j!k!}
R_{mj} R_{nk} \delta_{m-j\;n-k}.  \label{eq:Q_(mn)_expansion_gen}
\end{eqnarray}

For the particular case $\alpha =0$ and $\beta =-i\kappa $, $\kappa $ real,
we have
\begin{eqnarray}
&&Q_{mn}=\left( -i\right) ^{\left| m-n\right| }\exp {\left( \frac{-\kappa
^{2}}{4}\right) }  \nonumber \\
&&\times \left[ \sum_{j=0}^{\min \left( m,n\right) }\frac{\left( -1\right)
^{j}\kappa ^{\left| m-n\right| +2j}R{\left( \min \left( m,n\right) ,j\right)
}R{\left( \max \left( m,n\right) ,\left| m-n\right| +j\right) }}{2^{\left|
m-n\right| /2+j}j!\left( \left| m-n\right| +j\right) !}\right] .  \nonumber
\\
&&  \label{eq:Q_(mn)_expansion_alpha=0}
\end{eqnarray}

The apparently more general integral
\begin{equation}
\tilde{Q}_{mn}\left( \alpha _{1},\alpha _{2},\beta \right) =\int_{-\infty}^{\infty }\psi _{m}(x+\alpha _{1})^{*}\psi _{n}(x+\alpha _{2})\,e^{\beta x}\,dx  \label{eq:Q_(mn)_inte_2}
\end{equation}
can be reduced to the more suitable form (\ref{eq:Q_(mn)_inte}) by means of
the change of variables $q=x+\alpha _{2}$ that leads to $\tilde{Q}_{mn}\left( \alpha _{1},\alpha _{2},\beta \right) =e^{-\beta \alpha_{2}}Q_{mn}(\alpha _{1}-\alpha _{2},\beta )$.

\section{Expectation values}

Consider a wave packet for the potential moving with velocity $v_1$ and acceleration $a_1$:
\begin{equation}
\Psi(x,t) = \sum_{l=0}^\infty c_l \Psi_n(x,t) 
\end{equation}
where
\begin{equation}
\Psi_n(x,t) = \phi_n(\xi )  \exp \left(i \sigma(x,t)\right)  \ .
\end{equation}
and $\xi = x -x_1 -v_1 (t-t_1) - \frac{a_1}{2} (t-t_1)^2$. 

With easy manipulations we have
\begin{equation}
\begin{split}
\langle \Psi | \hat{x} | \Psi \rangle 
&= \Delta + \sqrt{\frac{\hbar}{2 m \omega}} \sum_{l} \left[ c_{l-1}^\star c_l \sqrt{l} e^{- i \omega (t-t_1)} +  \sqrt{l+1} c^\star_{l+1} c_l e^{i \omega (t-t_1)} \right]
\label{eq_x}
\end{split}
\end{equation}
and
\begin{equation}
\begin{split}
\langle \Psi | \hat{x}^2 | \Psi \rangle 
&= \sum_l  \Delta \sqrt{\frac{2\hbar }{m \omega }} \left[  \sqrt{l} c_l c_{l-1}^\star  e^{-i \omega (t-t_1)}
+  \sqrt{l+1} c_l c_{l+1}^\star e^{i \omega  (t-t_1)} \right] \\
&+ \sum_l  \frac{\hbar}{2 m \omega}
\left[ \sqrt{l-1} \sqrt{l}   c_l c_{l-2}^\star e^{-2 i \omega  (t-t_1)}
+ \sqrt{l+1} \sqrt{l+2} c_l c^\star_{l+2} e^{2 i \omega (t-t_1)} 
+ c_l c^\star_l (2 l +1 ) \right] \\
&+ \Delta^2
\end{split}
\label{eq_x2}
\end{equation}
where we have introduced for convenience $\Delta \equiv x_1 + v_1 (t-t_1) + \frac{a_1}{2} (t-t_1)^2$.

\begin{equation}
\begin{split}
(\Delta x)^2 &= \frac{\hbar}{2 m \omega} \sum_l  
\left[ \sqrt{l-1} \sqrt{l}   c_l c_{l-2}^\star e^{-2 i \omega  (t-t_1)}
+ \sqrt{l+1} \sqrt{l+2} c_l c^\star_{l+2} e^{2 i \omega (t-t_1)} 
+ c_l c^\star_l (2 l +1 ) \right]   \\
&- \frac{\hbar}{2 m \omega} \left(  \sum_{l} \left[ c_{l-1}^\star c_l \sqrt{l} e^{- i \omega (t-t_1)} +  \sqrt{l+1} c^\star_{l+1} c_l e^{i \omega (t-t_1)} \right] \right)^2
\end{split}
\label{eq_Delta_x}
\end{equation}

\section{Classical sho with two sudden changes of velocity and/or acceleration}

We consider a classical particle  in a simple harmonic oscillator that is at rest at $t < t_1$. We can choose $t_1=0$ for convenience.
At $t=t_1$ the potential starts suddenly to move with velocity $v$ and acceleration $a$ until, at $t=t_2$, it suddenly stops. We want to know the final energy
of the particle (at $t \leq t_2$).

We assume that at $t=0^{-}$ the particle has energy $E_1$ and is at the position $x_0$; its velocity is then $v_1 = \pm \sqrt{2/m (E_1-m \omega^2 x_0^2/2)}$. At $t>0^+$ the particle is subject to  a force $F=-m \omega^2 x - m a$ and the solution to the equations of motion are
\begin{equation}
\begin{split}
x_2(t) &= - \frac{a}{\omega^2} + A_2 \cos (\omega t +\varphi_2) \\
v_2(t) &= - A_2 \omega \sin (\omega t +\varphi_2)
\end{split}
\end{equation}

The amplitude and phase in this solution are obtained by enforcing the initial conditions:
\begin{equation}
\begin{split}
x_2(0^{+}) &= x_0 \\
v_2(0^{+}) &= -v + v_1 \\
\end{split}
\end{equation}

At $t=t_2$  the potential suddenly stops: in this case we can write the solution for $t \geq t_2$ as~\footnote{Notice that $ v t_2 + \frac{1}{2} a t_2^2$ is the position of the minimum of the potential at $t\geq t_2$}:
\begin{equation}
\begin{split}
x_3(t) &= v t_2 + \frac{1}{2} a t_2^2 + A3 \cos (\omega t +\varphi_3) \\
v_3(t) &= -A_3 \omega \sin(\omega t +\varphi_3) \\
\end{split}
\end{equation}

The amplitude and phase in this solution are obtained by enforcing the initial conditions:
\begin{equation}
\begin{split}
x_3(t_2^{+}) &= v t_2 + \frac{1}{2} a t_2^2 + x_2(t_2) \\
v_3(t_2^{+}) &= v + a t_2 + v_2(t_2)\\
\end{split}
\end{equation}

The final energy is then
\begin{equation}
E_3 = \frac{1}{2} m \omega^2 A_3^2 
\end{equation}

To facilitate the comparison with the quantum result we parametrize the initial classical energy as
$E_1 = \epsilon \frac{\hbar \omega}{2}$ and consider the ratio $E_3/E_1$. It is convenient to express
the results in dimensionless form using the parameters $\kappa$ and $\lambda$ and defining
$y \equiv \sqrt{m\omega x_0/\hbar }$; we find
\begin{equation}
\begin{split}
\frac{E_3}{E_1} &= \frac{1}{\epsilon } \left[ \epsilon +2 \left(\kappa ^2+\lambda  (y+\lambda )\right)+2 \kappa \lambda  \tau +\lambda ^2 \tau ^2 \right. \\
&- \left. 2 \left(\kappa ^2+\lambda 
   (y+\lambda )+\kappa  \lambda  \tau \right) \cos (\tau )-2 (y \kappa
   +\lambda  (y+\lambda ) \tau ) \sin (\tau ) \right. \\
&\pm \left. 2 \sqrt{-y^2+\epsilon }
   (\kappa -(\kappa +\lambda  \tau ) \cos (\tau )+\lambda  \sin (\tau
   )) \right]
\end{split}
\end{equation}

Notice that there are two possibilities ( due to the sign $\pm$) and that the energy depends on the initial position of the particle at time $t_1$. Therefore we need to average over the initial position 
\begin{equation}
\begin{split}
\overline{\frac{E_3}{E_1}} &\equiv \frac{1}{2\sqrt{\epsilon}} \int_{-\sqrt{\epsilon}}^{\sqrt{\epsilon}}
\frac{E_3}{E_1}  dy \\
&= \frac{1}{2\epsilon} \left[ 2 \epsilon  \pm \pi   \sqrt{\epsilon } \kappa + 4 \kappa ^2+4
   \kappa  \lambda  \tau +2 \lambda ^2 \left(2+\tau ^2\right) \right. \\
&- \left. \left(4
   \lambda ^2+\left( \pm \pi  \sqrt{\epsilon }+4 \kappa \right)
   (\kappa +\lambda  \tau )\right) \cos (\tau )+\lambda  \left(\pm \pi 
   \sqrt{\epsilon }-4 \lambda  \tau \right) \sin (\tau ) \right]
\end{split}
\label{eq_sho_cl}
\end{equation}

Since there two different expressions and correspond to equally probable situations, we still need to take the statistical average  of the two
\begin{equation}
\begin{split}
\left. \overline{\frac{E_3}{E_1}} \right|_{s.a.} &\equiv \frac{1}{\epsilon} \left[ \epsilon +2 \kappa^2+2 \kappa  \lambda  \tau +\lambda^2 \left(2+\tau ^2\right)
\right. \\
&- \left. 2 \left(\kappa ^2+\lambda^2+\kappa  \lambda  \tau \right) \cos (\tau )-2 \lambda ^2 \tau  \sin (\tau ) \right]
\end{split}
\end{equation}

\end{document}